\documentclass[epsfig,onecolumn,10pt]{article}
\usepackage[]{graphicx}
\begin{document}
\newif\ifincludefigures
\newif\ifnofigures

\includefigurestrue
\includefiguresfalse


\thispagestyle{empty}


\vspace*{\baselineskip}
\vspace*{\baselineskip}

{\noindent\Large \bf Aerogel keystones:  extraction of complete hypervelocity impact events from aerogel collectors}
\ \\
\ \\
\textbf{\large \bf Andrew J. Westphal$^1$, Christopher Snead$^1$, Anna Butterworth$^1$, Giles A. Graham$^2$, John P. Bradley$^2$, Sa\v sa Bajt$^3$, Patrick G. Grant$^4$, Graham Bench$^4$, Sean Brennan$^5$, Piero Pianetta$^5$}\\

\noindent$^1$ Space Sciences Laboratory, University of California at Berkeley, Berkeley, CA 94720 USA \\
\noindent$^2$ Institute for Geophysics and Planetary Physics, Lawrence Livermore National Laboratory, Livermore, CA 94035 USA \\
\noindent$^3$ Lawrence Livermore National Laboratory, Livermore, CA 94035 USA \\
\noindent$^4$ Center for Accelerator Mass Spectrometry,  Lawrence Livermore National Laboratory, Livermore, CA 94035 USA  \\
\noindent$^5$ Stanford Synchrotron Radiation Laboratory, 2575 Sand Hill Road, Menlo Park, CA 94550 USA  \\

\noindent\dotfill \\
\noindent\textbf{
In January 2006, the Stardust mission will return the first samples from a 
solid solar-system body since Apollo, and the first samples of 
contemporary interstellar dust ever collected.  Although sophisticated 
laboratory instruments exist for the analysis of Stardust samples, 
techniques for the recovery of particles and particle residues from aerogel 
collectors remain primitive.  Here we describe our recent progress in 
developing techniques for extracting small volumes of aerogel, which
we have called ``keystones,'' which completely contain particle impacts but
minimize the damage to the surrounding aerogel collector.  These keystones
can be fixed to custom-designed micromachined silicon fixtures (so-called
 ``microforklifts''). In this configuration the samples are self-supporting, 
which can be advantageous  in situations in which interference from a
 supporting substrate is undesirable.  The keystones may also be extracted 
and placed onto a substrate without a fixture. We have also demonstrated 
the capability of homologously crushing these unmounted keystones for 
analysis techniques which demand flat samples.
}

\clearpage

\ \\
{\bf \Large Introduction: Stardust science goals and required analytical techniques}
\ \\


In this first paper in an anticipated technical series from the 
Bay Area Particle Analysis Consortium (BayPAC)\footnote{Building on our existing extraction
development effort \cite{westphal}, and leveraging the extensive analytical capabilities available in the San Francisco Bay Area, we have formed a consortium of National Laboratories and Universities:  Lawrence Livermore National Laboratory (LLNL), Space Sciences Laboratory (SSL) at the University of California at Berkeley, the Advanced Light Source (ALS) at the Lawrence Berkeley National Laboratory (LBNL) and the Stanford Synchrotron Radiation Laboratory (SSRL) at the Stanford Linear Accelerator Center (SLAC), with a focus on preparation for the return of Stardust samples.}, 
we describe our recent efforts in the development of advanced techniques for the extraction
of hypervelocity impacts in silica aerogel collectors, and subsequent preparation of these extracted
residues for detailed laboratory analysis.  In subsequent papers we will discuss the analysis of
keystone-extracted impacts using a variety of analytical techniques.


The Stardust mission, which was launched in 1999, is expected to collect cometary dust
from the coma of comet Wild-2 in January 2004.  The aerogel collector will be returned to earth
for laboratory analysis in January 2006.  If successful, Stardust will return the first samples from a 
planetary body since Apollo.   The importance of this mission to planetary science
cannot be overstated.   Wild-2 is a long-period comet which was placed into a short-period
orbit, with perihelion just inside the orbit of Mars,   through a fortuitous encounter 
with Jupiter in 1974.  Wild-2 is probably composed of extremely primitive material 
that has suffered very little alteration since its accretion into a solid body $\sim 4.6$ Gy ago; 
indeed, it is not impossible that it is composed  of nearly pristine interstellar material.
By the time of the Stardust encounter, Wild-2 will have orbited the Sun only $\sim 5$ times
since the beginning of its residence in the inner solar system.


Since Stardust will return the first samples known with certainty to be cometary material, the
zeroth-order questions for Stardust will be:  

\begin{itemize}%
\item Have we seen this type of material before?
\item Is it similar to meteorites, micrometeorites, or interplanetary dust particles (IDPs)?
\item Is it solar system material, presolar material, or both?
\item Are there organics and if so what are they (aliphatic, aromatic, N-containing)?
\end{itemize}

The broader scientific questions that Stardust will eventually address are:

\begin{itemize}%
\item To what degree was material from the inner solar system mixed with that of the outer solar system in the presolar nebula?  Are the reservoirs of material for the two regions distinct?
\item How is cometary material related to interstellar dust?   What is the origin of crystalline silicates
in cometary material, given that interstellar silicates appear to be amorphous?
\item What is the origin of organics in cometary dust?  Are the organics interstellar in origin,
or were they formed in the presolar nebula?
\end{itemize}

The analyses directed at each of these questions will require overlapping and complementary analytical techniques.Ê For example, to address the first two zeroth-order questions listed above, detailed mineralogical/petrological analysis using analytical electron microscopy will likely be required.ÊÊÊ But measurements of the elemental compositions using for example Proton-Induced X-ray Emission (PIXE) or X-Ray Fluorescence (XRF) will also be required.Ê The third zeroth-order question will require ion microprobe analyses to measure the isotopic compositions of individual grains.Ê Identification of organics will require sophisticated analytical techniques (e.g synchrotron infrared spectroscopy and Raman spectroscopy) possibly coupled with isotopic analyses. Given the spectrum of analytical procedures sample preparation procedures must be as universal as possible in order to take full advantage of the full complement of analytical techniques.Ê As we developed particle extraction techniques, we kept this requirement fully in mind.

The Stardust collector consists of 130 aerogel tiles, with dimensions  42mm $\times$ 21mm  on the collecting face,  30mm thickness,  and density  in a gradient 
from $\sim$10 mg cm$^{-3}$ at the collecting surface to $\sim$50 mg cm$^{-3}$ at the bottom of the collector.
The goal of Stardust is to collect more than $1000$ particles at least $15{\mu}$m in size, and many millions of smaller particles.    Because of large uncertainties in the modelling of dust in 
the cometary coma, this estimate of the  statistics is subject to order-of-magnitude uncertainties.
Large deviations in these statistics in either direction could pose a problem for Stardust analysis.
A yield much smaller than expected could compromise the scientific yield of Stardust.  
A yield much larger than expected
could compromise the structural integrity of the collector and could make {\em in situ}
characterization of individual particles difficult or impossible.    We address this possibility
further later in this paper.

\ \\{\bf \large Hypervelocity impacts of friable chondritic particles in aerogel } \ \\

\begin{figure}
\begin{center}
\rotatebox{0}{\resizebox{!}{7cm}{\includegraphics*{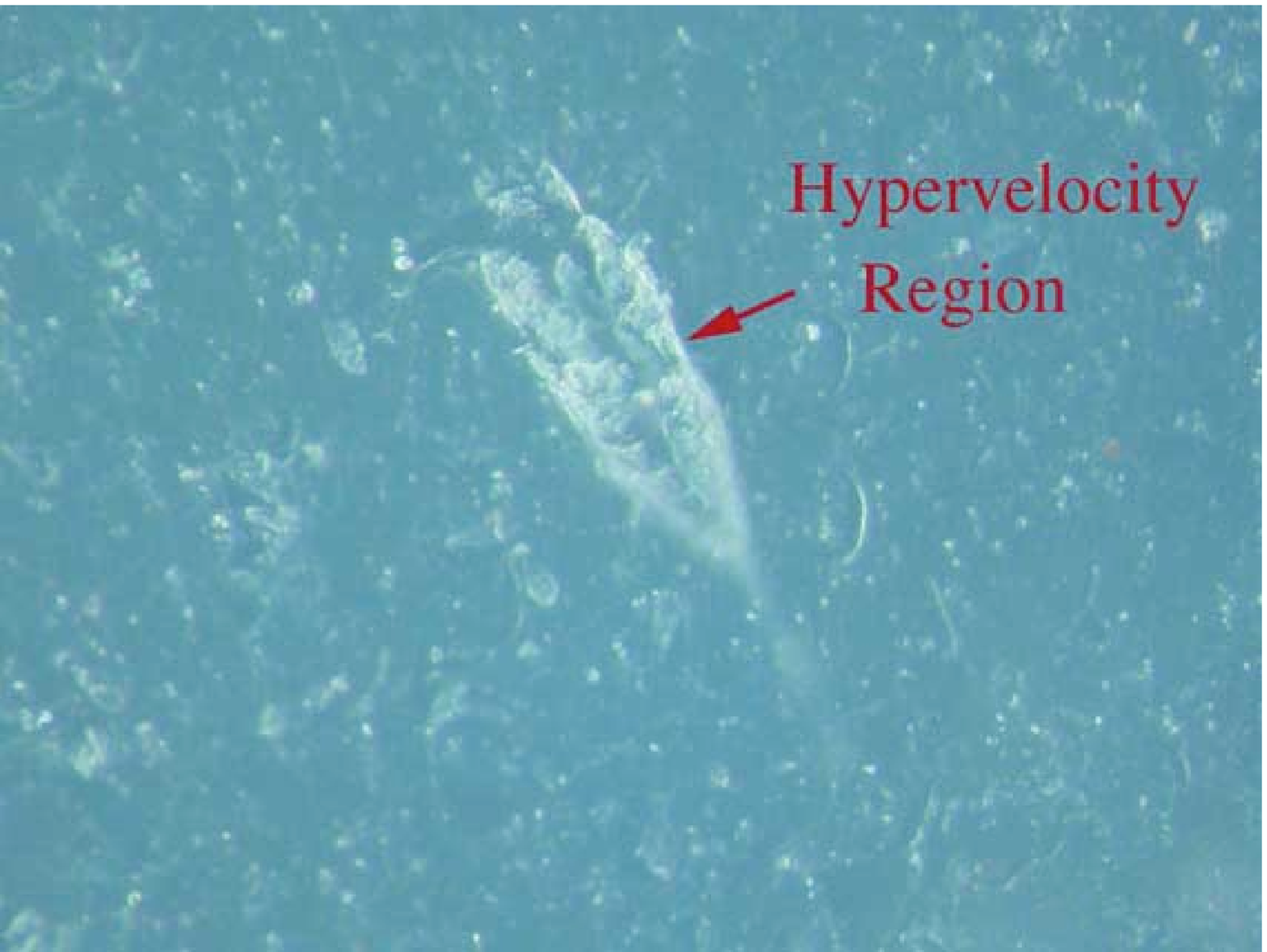}}}
\rotatebox{0}{\resizebox{!}{7cm}{\includegraphics*{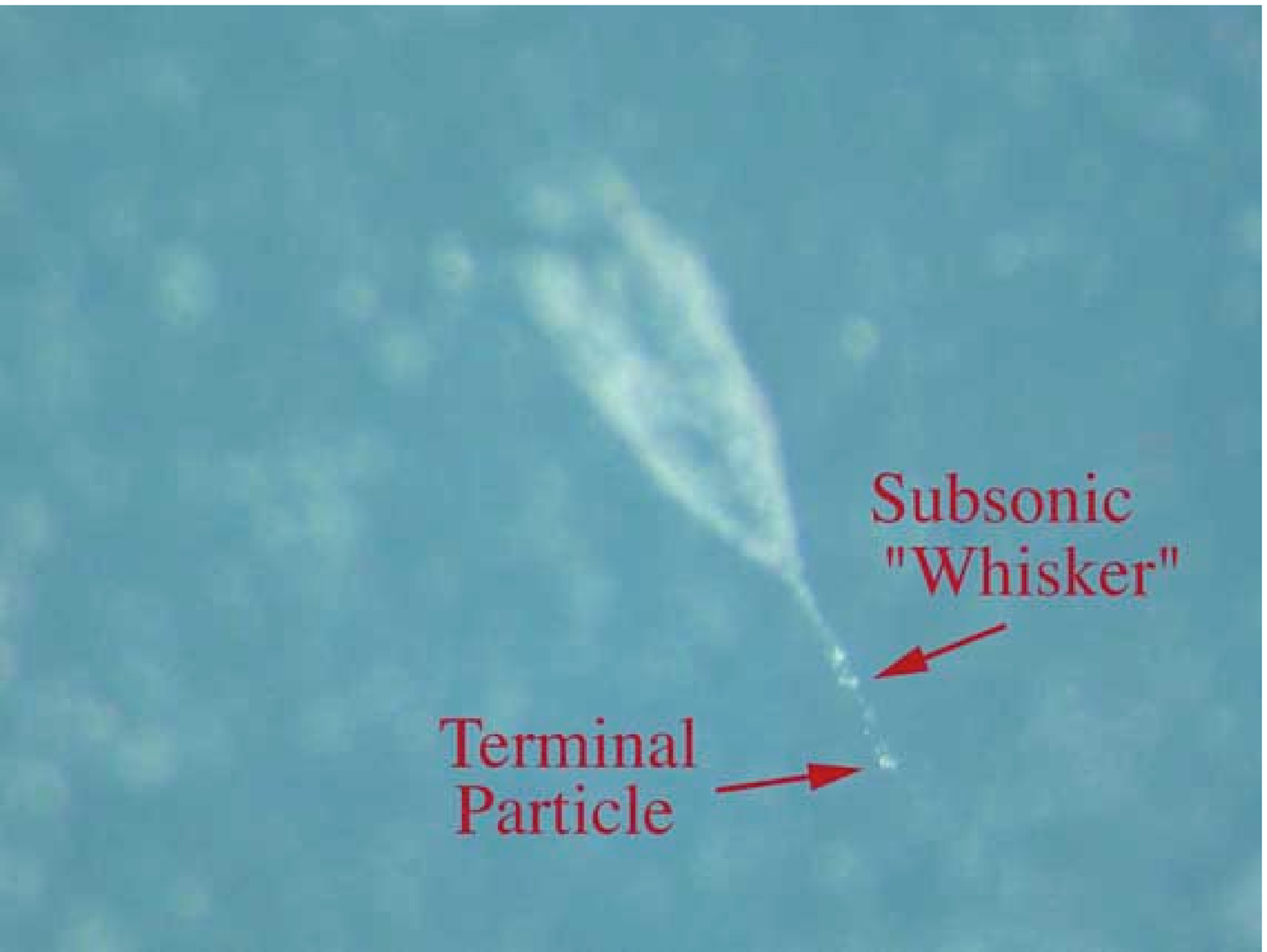}}}
\caption[carrot-track]{Anatomy of a hypervelocity particle impact track in aerogel. } \label{carrot-track}
\end{center}
\end{figure}

Particles with speeds of a few km sec$^{-1}$  or greater typically produce 
characteristic carrot-shaped tracks when they stop in aerogel collectors (Fig. \ref{carrot-track}).  
Particles with initial sizes $\sim10\mu$m in size typically produce tracks that are several hundred
microns long.  Near the entry point, the particle moves hypersonically and produces a strong, cylindrical shock, forming a tubular cavity (the ``carrot'').  
As the shock expands into the aerogel collector, 
it weakens and finally stops when  the shock pressure no longer exceeds the 
crushing pressure of the aerogel.    The evolution of this shock is similar in 
some respects to the evolution of supernova shocks in the interstellar medium.
If the particle is sufficiently fragile, it is disrupted into fragments during the hypervelocity
phase.   As each surviving particle slows,  the shock near the particle weakens until it is either  too weak to
 crush the aerogel or becomes an acoustic wave.   In this region, one or more thin ``whisker'' tracks are
produced.  Eventually, the surviving particles stop when their ram pressure no longer 
exceeds the crushing pressure of the aerogel.   Dominguez and Westphal\cite{dominguez} have developed a detailed model incorporating these physics that accurately predicts the shape and size of these impact tracks for non-fragmenting projectiles. This simple picture is 
complicated by three effects:  ablation, which can be severe for particles
with a large volatile component; fragmentation, which we discuss below; and accretion, in
which molten and compressed aerogel accretes onto the stopping particle.  Aerogel accretion
has been reported to at least partially protect the projectiles from damage \cite{barrett}, although
this effect has not been studied extensively.   Dominguez and Westphal\cite{dominguez} speculate
that the large dispersion in observed range is due to repeated cycles of gradual accretion and abrupt, random shedding
of heated aerogel by stopping hypervelocity particles.

\noindent{\bf }

\begin{figure}
\begin{center}
\rotatebox{0}{\resizebox{!}{7cm}{\includegraphics*{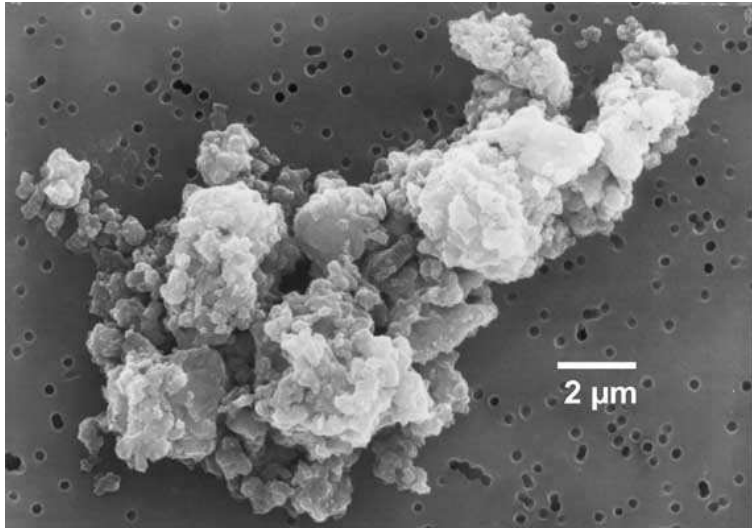}}}
\caption[idpimage]{Image of a typical anhydrous cluster IDP. } \label{idpimage}
\end{center}
\end{figure}

In our approach to the development of particle extraction techniques, we have assumed
 that a reasonable analog for cometary particles collected from the Wild-2 coma
will be anhydrous Interplanetary Dust Particles (IDPs), which are generally suspected
of being cometary in origin\cite{brownlee}.  These particles, which have been collected in the stratopshere for many years,  are fractal-like aggregates of small grains, but often include large, relatively refractory grains (Fig. \ref{idpimage}).  
These particles (or their  artificially-synthesized analogs) are so 
fragile that they do not survive the strong shocks
present inside hypervelocity gas guns, so we have no empirical information on survivability
of such particles when captured by aerogel.    However, we have assumed that even the
mild shock pressures experienced by the particles during capture in aerogel are very likely
to disaggregate these fragile particles into their components.  Since the range of particles in
aerogel is dependent mostly on the particle size, the small components  stop quickly, while
any large and robust particles present  have the longest range and  form the track termini.  
At first inspection under an optical microscope, the resulting track can
give the mistaken impression the entire particle was captured intact, or at least
disintegrated into only a few fragments, since the small  components are difficult to 
image and easy to overlook.   
Since the large, robust components are typically mineral grains, an analysis of the track which does not include the distributed material along the track
will be seriously biased towards robust minerals.  If only the terminal particles
from the Stardust samples are to be recovered and analyzed, we might conclude that comet Wild-2
is composed entirely of forsterite!

\noindent{\bf }

\begin{figure}
\begin{center}
\rotatebox{0}{\resizebox{!}{7cm}{\includegraphics*{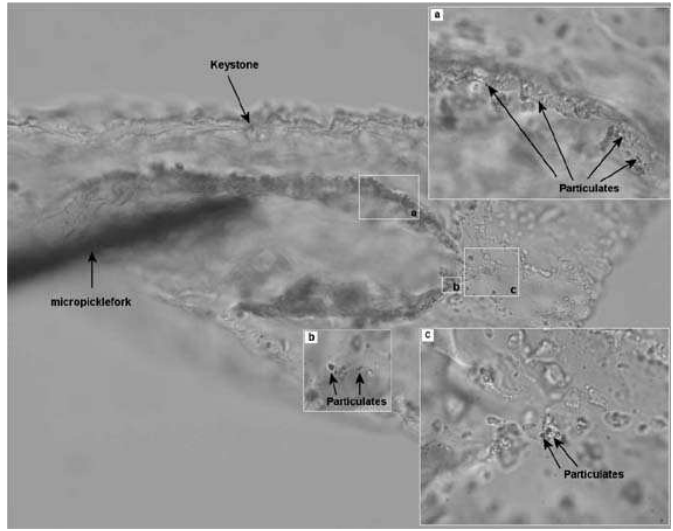}}}
\caption[calorimetry]{  A ``chondritic swarm'' impact event observed in transmitted light in an optical microscope.  In addition to the large terminal particle, fine-grained particle 
residue is distributed along the length of the impact track.} \label{optical-distribution}
\end{center}
\end{figure}

The assumption that captured chondritic particles readily shed small fragments on capture 
is at least partially confirmed by a population of impact events in the Orbital Debris Collection Experiment (ODCE) collector.  OCDE was deployed on the Russian space station  {\em Mir} for 18 months starting in March 1996.  Zolensky {\em et al.} have made a detailed examination of the impacts in this collector \cite{odce}.    A fortunate happenstance has produced a large number of parallel impacts with chondritic composition in the ODCE collector.  The working hypothesis
is that a relatively large chondritic ($\gg100{\mu}$m) particle, tentatively identified by 
Zolensky as a CV3 meteorite\cite{odce}, suffered a glancing collision with a component of the space station; 
this  collision was sufficiently strong to disrupt the particle into numerous fragments, but was
sufficiently peripheral that it produced very little ejecta from {\em Mir}.
These particle impacts have proven to be very useful as targets for the development of
extraction techniques, since their typical size ($10{\mu}$m or smaller) are probably
typical of the sizes of particles to be extracted from Stardust, and they are known to be chondritic.
Detailed examination of these impacts in an optical microscope (Fig. \ref{optical-distribution}) and using a nuclear microprobe (Graham {\em et al}, in preparation)
shows the presence of a  substantial amount of material distributed along the tracks of these impacts.  Since CV3 meteorites are probably much less friable than cometary material, we expect cometary impacts to be even more strongly biased away from large, intact terminal particles.  
We have speculated that cometary impacts may in fact resemble the descriptively-named 
hedgehog events observed in ODCE \cite{odce}.  These events resemble small explosions in the aerogel, perhaps consistent with flash-vaporization of volatile materials.   Although particles are observed in hedgehogs, usually at terminus of radial tracks, no distinct terminal particle is usually present.

We conclude that a conservative approach to the preparation for Stardust analysis is to assume
that much --- perhaps most ---  of the cometary material will be distributed along impact tracks,
and that  if one neglects this material one may introduce serious systematic
biases into the analysis.  Thus, the entire impact event should be recovered if possible.  It
may also be important to disturb the remainder of the collector as little as possible.  These
requirements have driven our development effort.

\ \\{\bf \large End-to-end Stardust analysis}\ \\

\begin{figure}
\begin{center}
\begin{minipage}[t]{6.2in}
\leavevmode 
\rotatebox{0}{\resizebox{!}{20cm}{\includegraphics*{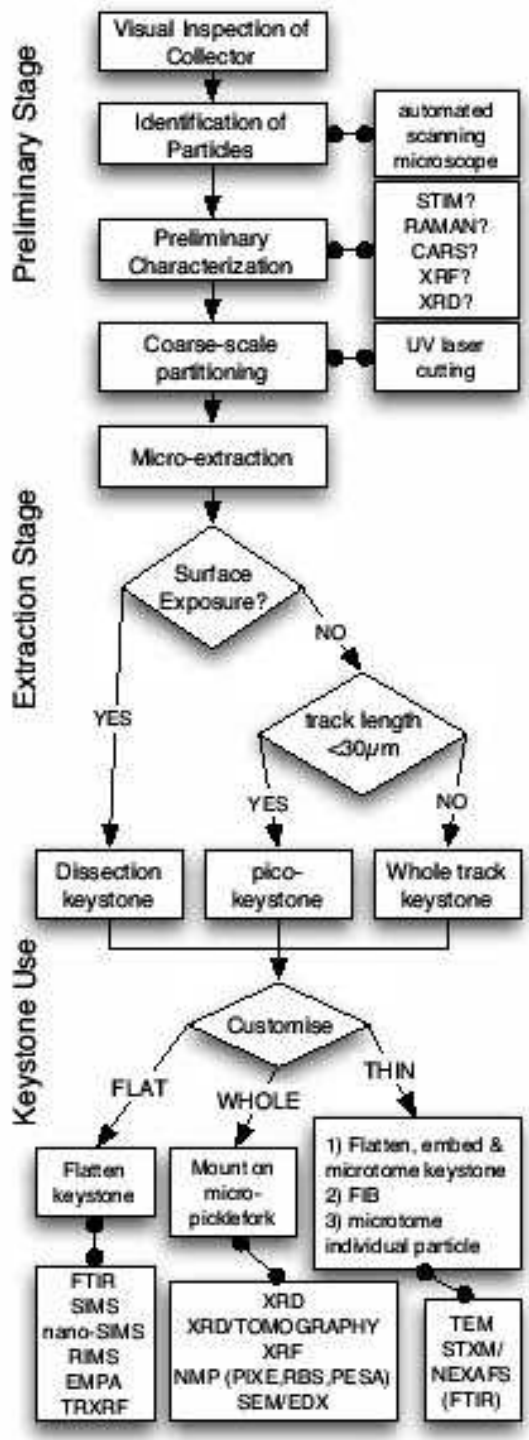}}}
\end{minipage}
\caption[flowchart]{Draft flowchart for Stardust analysis.} \label{flowchart}
\end{center}
\end{figure}

In Fig. \ref{flowchart} we show a draft 
flowchart for the analysis of Stardust samples, from
the initial examination of collectors and identification of particle impacts through
detailed measurements in laboratory instruments.


The protocol for initial examination
of collectors will be developed by the JSC curatorial facility.    Tools for initial examination
are well-developed.  Westphal {\em et al.} have previously
demonstrated that a automated scanning microscope can be used to identify particle
impacts in aerogel collectors \cite{westphal}.  Whether or not automated scanning will be required
for the Stardust cometary collector has not yet been established.   Other tools that may be useful for
 distribution of Stardust samples exist now:  for example, 
large-scale cutting of aerogel cells into smaller pieces using a pulsed 
UV laser has been demonstrated by Graham {\em et al}\cite{laser}.    Preliminary {\em in situ} characterization of particles could be done by 
Raman microscopy\cite{raman-refs} or Synchrotron X-ray Fluorescence\cite{sxrf,xrf-ref}.

In 2000,  Zolensky {\em et al.} \cite{smallisbeautiful} reviewed some of  the analytical 
techniques that will be available for Stardust.  Even in the short time since that review 
was written, spectacular progress has been made in many analytical techniques. 
For example, using the new Cameca nanoSIMS, Messenger {\em et al.} have
 identified presolar grains in IDPs\cite{messenger}.
The spatial resolution of the $^{17}$O maps of IDPs in their analyses was $\sim 100$ nm.   
Scanning Transmission X-ray Microscopy (STXM) can now routinely
map various bound states of CNO using NEXAFS  with better than 100 nm resolution \cite{stxm}. 
Further advances are anticipated in the next few years.   For example, George {\em et al.} have pointed
out that MEMS Force-Detected Nuclear Magnetic Resonance (MEMS FD-NMR) could enable
 NMR analysis on micron-scale particles\cite{FDNMR} because of the very weak scaling of its single-shot signal-to-noise ratio (SSSNR)  with sample size $m$
(SSSNR $\sim m^{3\over 2}$) in contrast with the strong scaling of conventional 
NMR (SSSNR $\sim m^6$).  (The two NMR techniques have comparable SSSNR at $m = 1$ mg.)  


But techniques for extraction of hypervelocity particle impacts from aerogel, and subsequent sample preparation for analysis in advanced laboratory instruments, remain primitive.  
In this paper,  we focus on the steps between {\em in situ} particle identification and analysis --- the extraction of impact residues and preparation of extracted samples for analysis.

\ \\{\bf \Large Fully-contained extraction of particle impacts in aerogel ``keystones''} \ \\

Precision cutting of aerogel on the micron scale is a difficult challenge\cite{westphal}.   To ``machine'' 
aerogel, we used borosilicate glass microneedles, manufactured in our laboratory 
using a commercially-available micropipette puller. 
We mounted the needles on commercially-available 3-axis micromanipulators (Sutter
Instruments MP-285), which were robotically controlled by an independent computer (Sun Sparc 1).
The cutting action consists of repeated small axial poking motions of the aerogel by
the microneedles.  A slice can be produced in the aerogel by poking repeatedly in a line, 
with increasing depths between each sequence of pokes.  We empirically determined
the parameters of the moves (speed, spacing between pokes, 
increase in depth between each sequence of pokes, etc.), and found that the optimal
parameters are different for aerogels from different manufacturers. 

\begin{figure}
\begin{center}
\rotatebox{0}{\resizebox{!}{4cm}{\includegraphics*{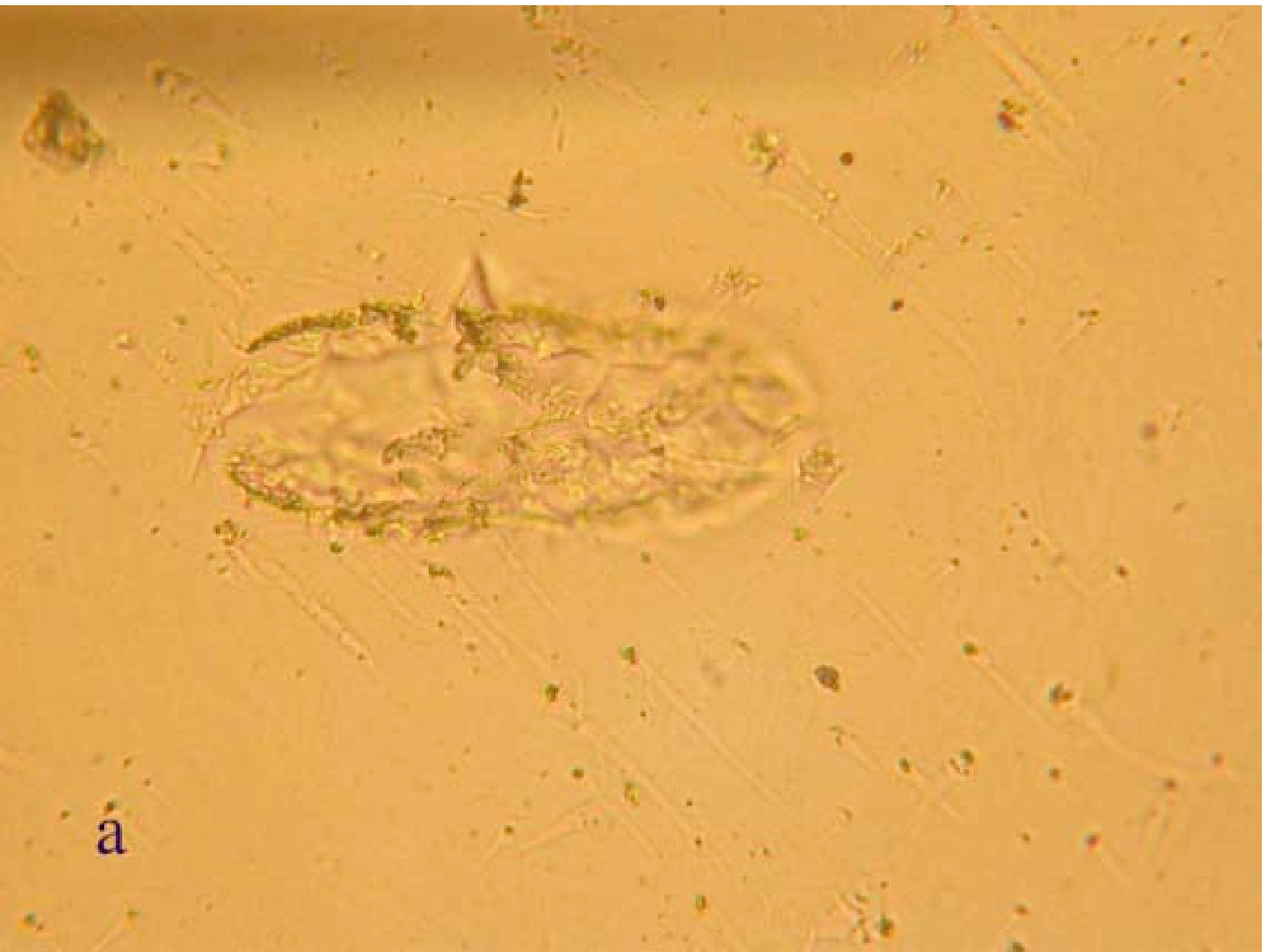}}}
\rotatebox{0}{\resizebox{!}{4cm}{\includegraphics*{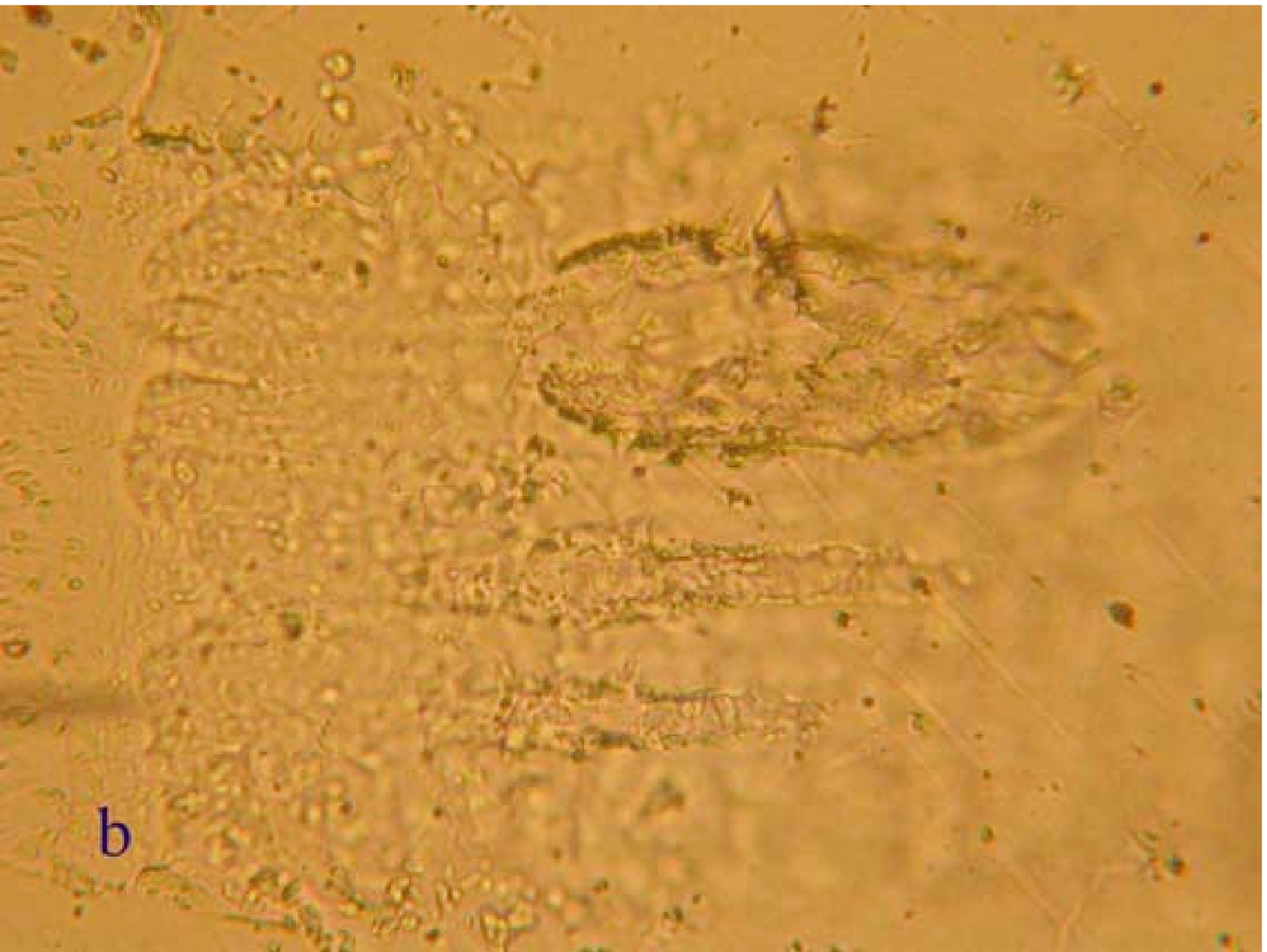}}}
\rotatebox{0}{\resizebox{!}{4cm}{\includegraphics*{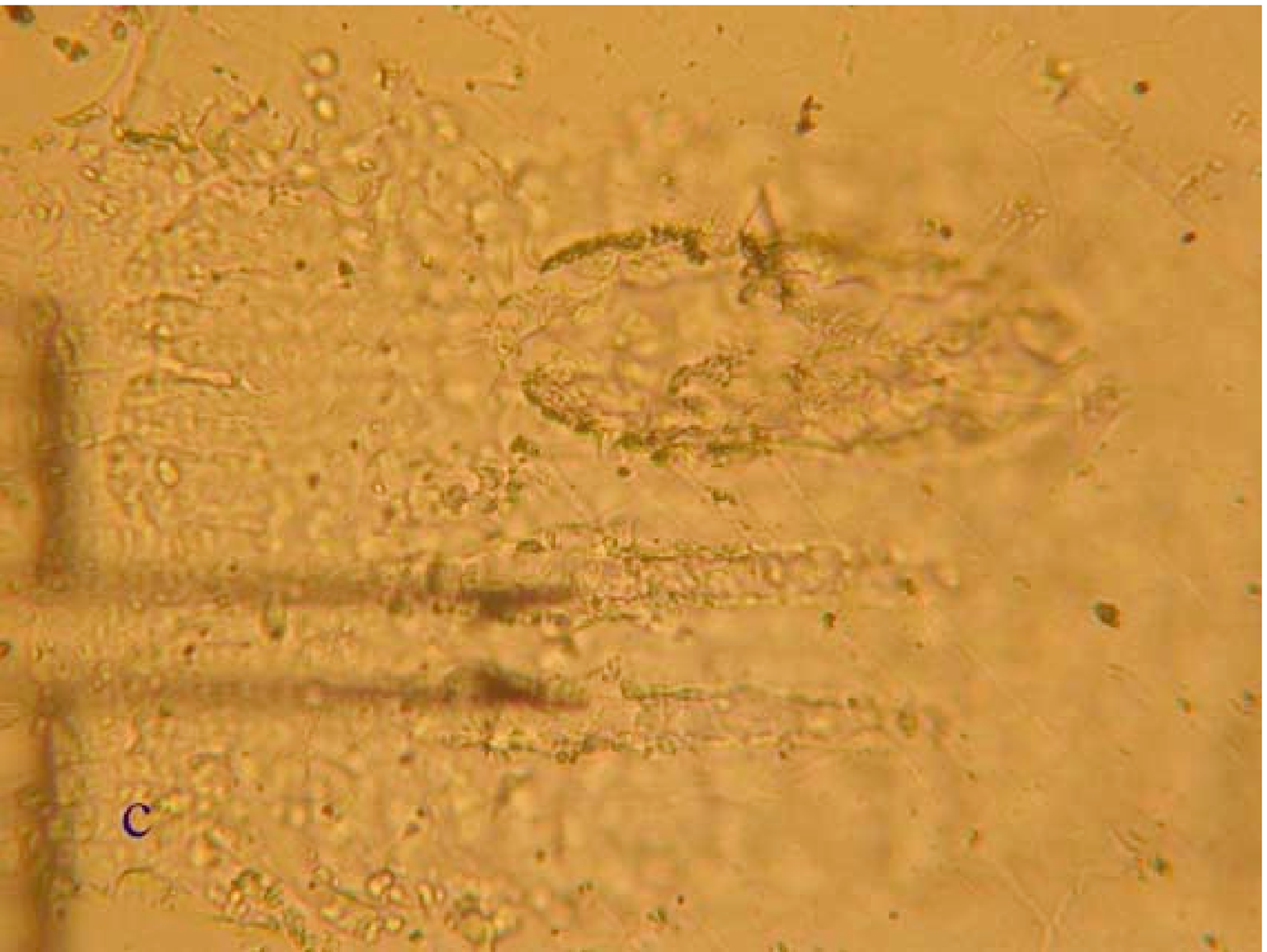}}}
\rotatebox{0}{\resizebox{!}{4cm}{\includegraphics*{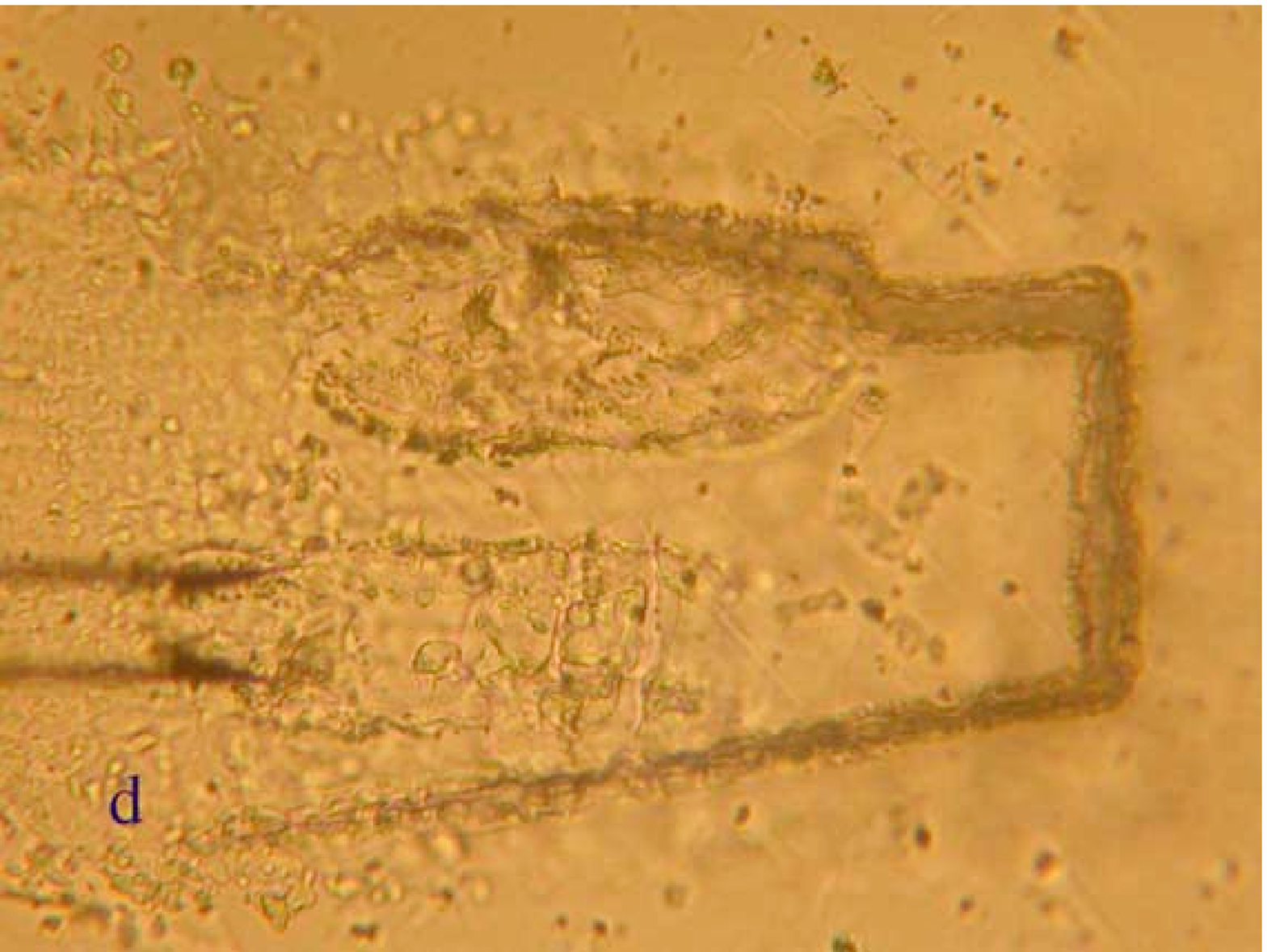}}}
\rotatebox{0}{\resizebox{!}{4cm}{\includegraphics*{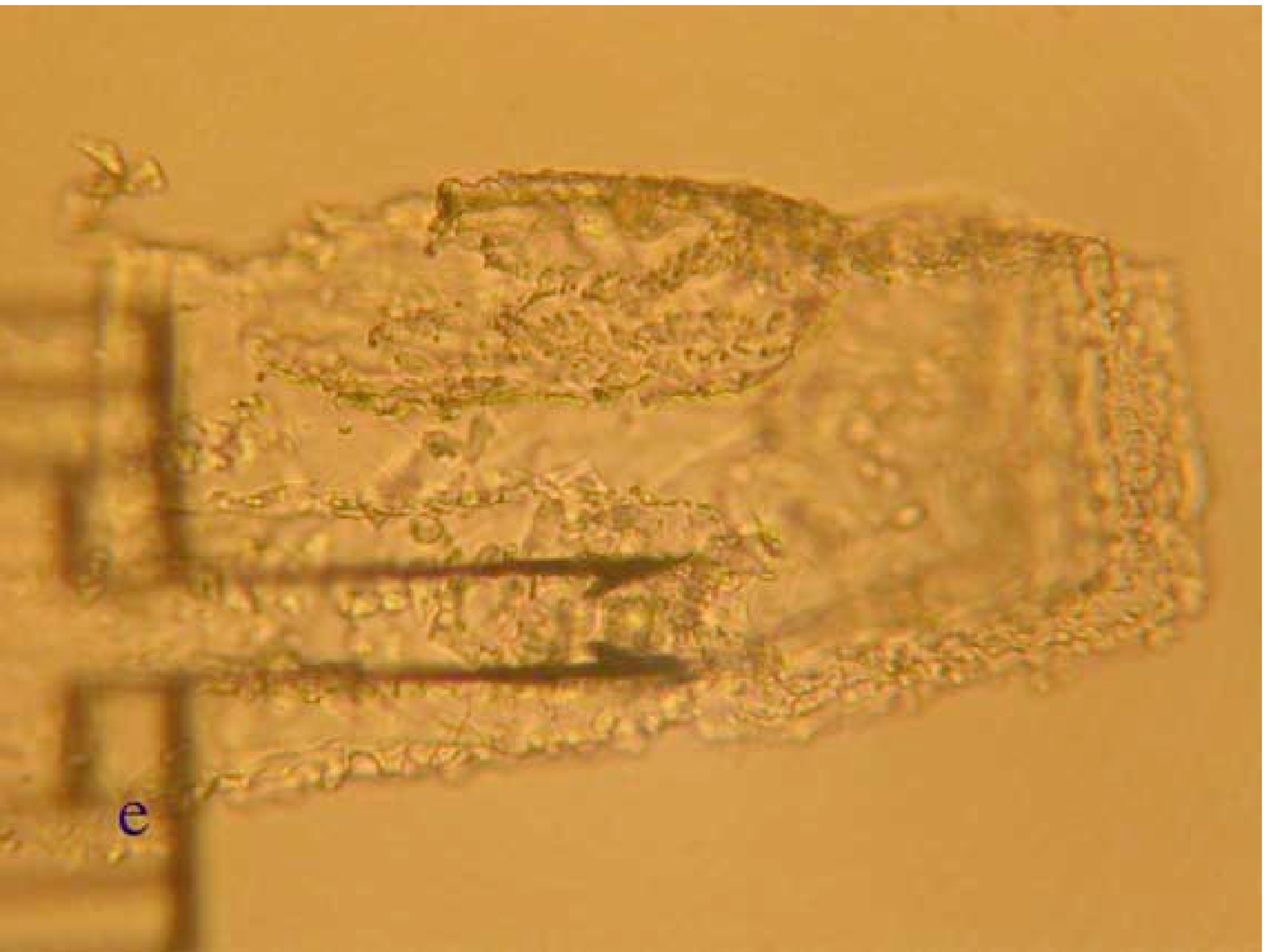}}}
\rotatebox{0}{\resizebox{!}{4cm}{\includegraphics*{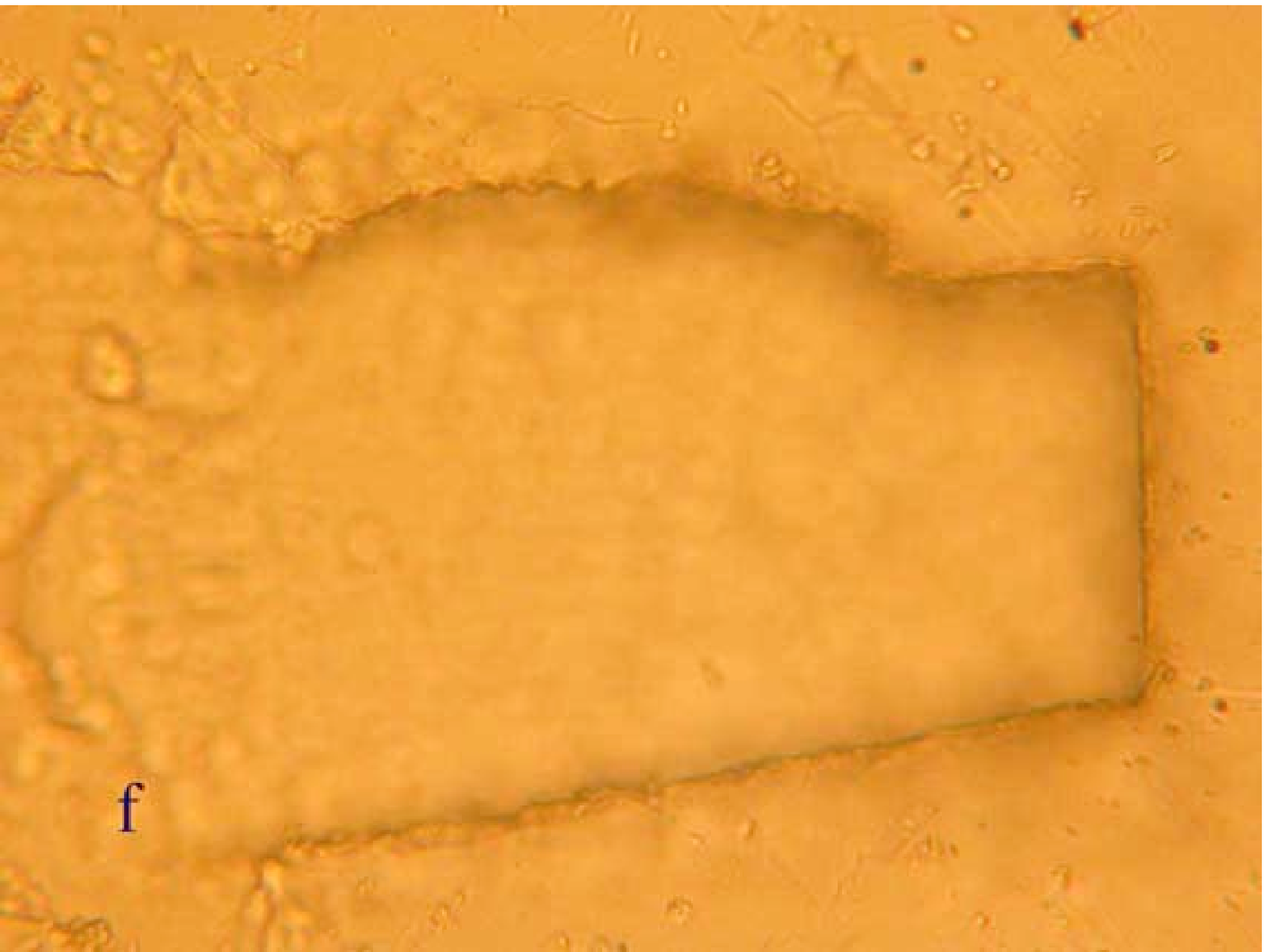}}}
\rotatebox{0}{\resizebox{!}{4cm}{\includegraphics*{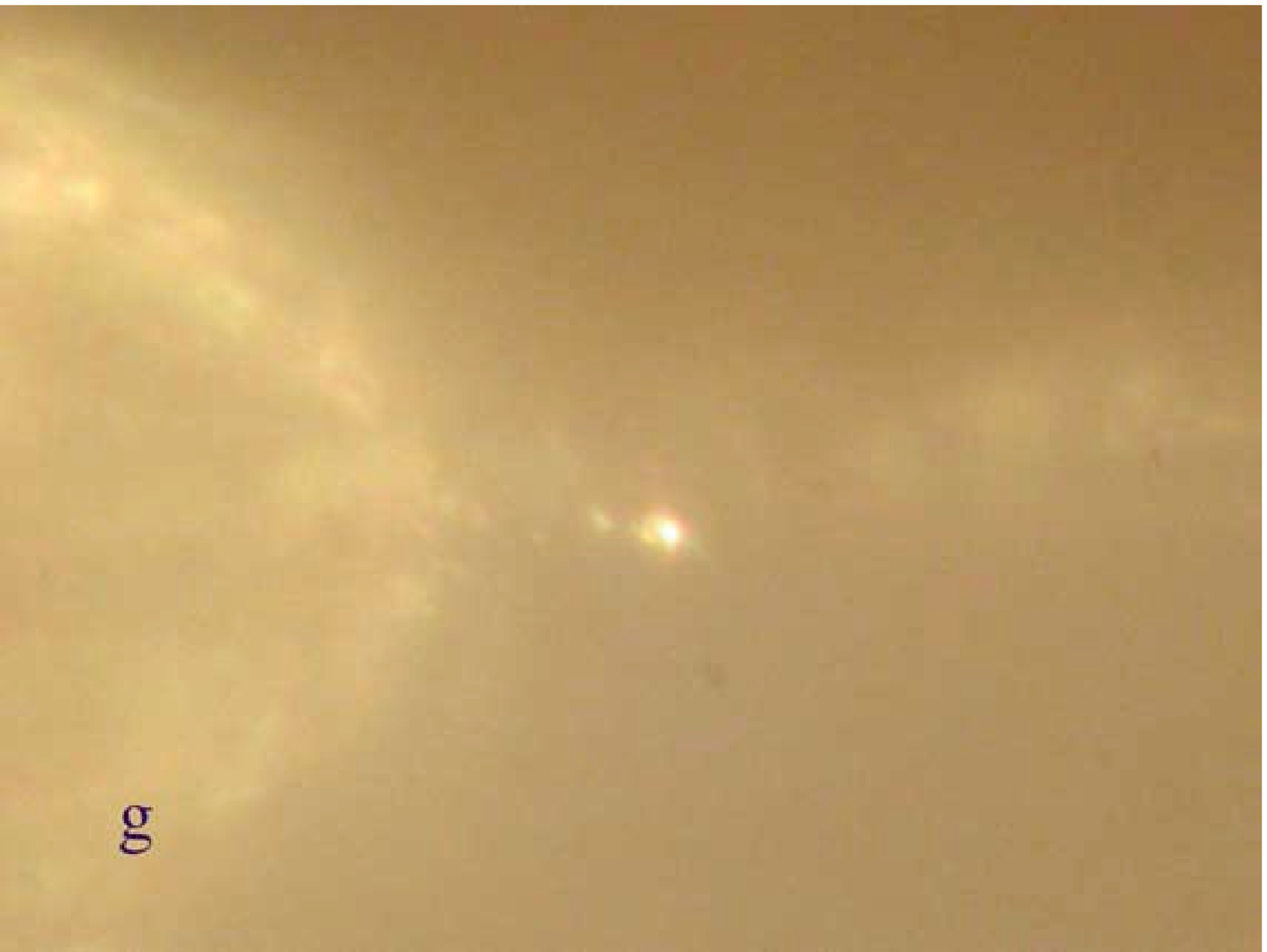}}}
\caption[extraction-schematic]{Procedure for extaction of a hypervelocity particle impact (121 $\mu$m depth) from an aerogel collector.  a) Impact in collector before extraction.  b) Undercut and microforklift handle emplacement.  c) Placement of microforklift.  d) Vertical contour cut.  e)  Extracted keystone (475 $\mu$m in longest dimension)  f) cavity in collector after keystone extraction g) Terminal particle in keystone (reflected light image) } \label{extraction-schematic}
\end{center}
\end{figure}

To extract a particle impact, we first align the impact along the x-axis of the microscope
stage.  We then undercut the impact using a straight needle mounted at shallow angle (typically
 27$^\circ$) to the horizontal.   (Fig. \ref{extraction-schematic}).  
Using the same needle, we cut two small tunnels that we use in a later
step for mounting the keystone in a micromachined fixture.  We then cut the event out of the collector
using a microneedle bent so that its tip is oriented normal to the collector surface.
Any shape can be defined for this cut.  Typically, we follow the particle impact track on one
side, leaving  $\sim10\mu$m of aerogel between the wall of the impact track
and the vertical cut.   This results in an aerogel sample (a ``keystone'' )
that completely contains and conserves the impact event.   The damage to the surrounding
aerogel is minimized.  

\noindent{\bf }

\begin{figure}
\begin{center}
\rotatebox{0}{\resizebox{!}{4cm}{\includegraphics*{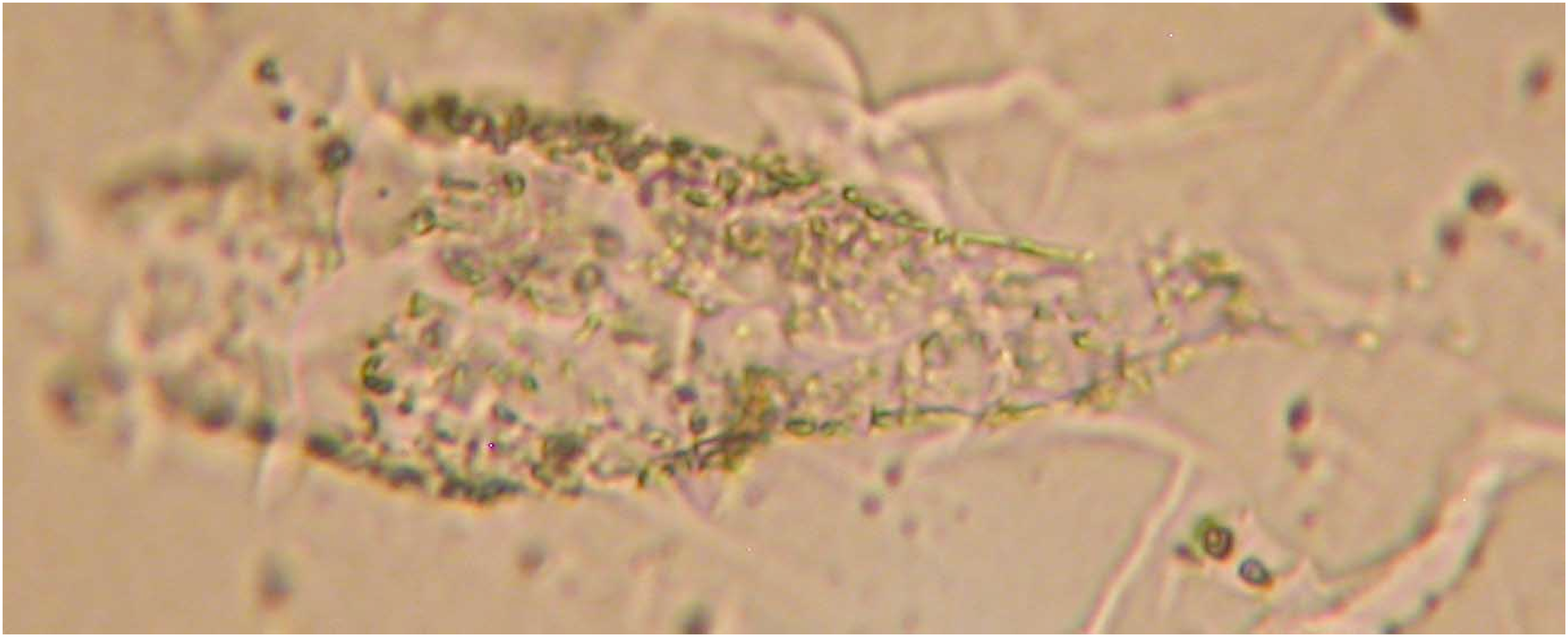}}}
\rotatebox{0}{\resizebox{!}{7cm}{\includegraphics*{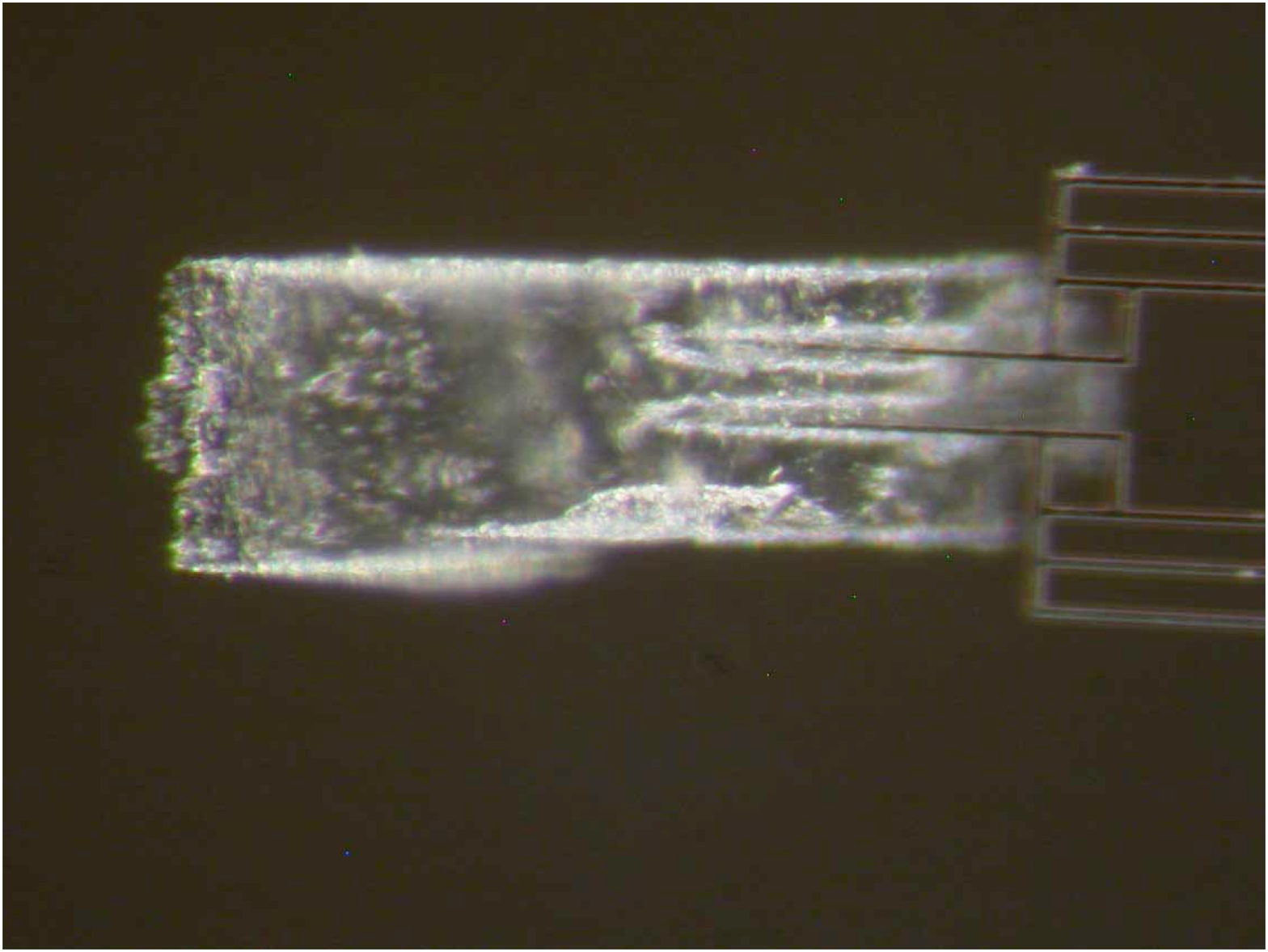}}}
\rotatebox{0}{\resizebox{!}{7cm}{\includegraphics*{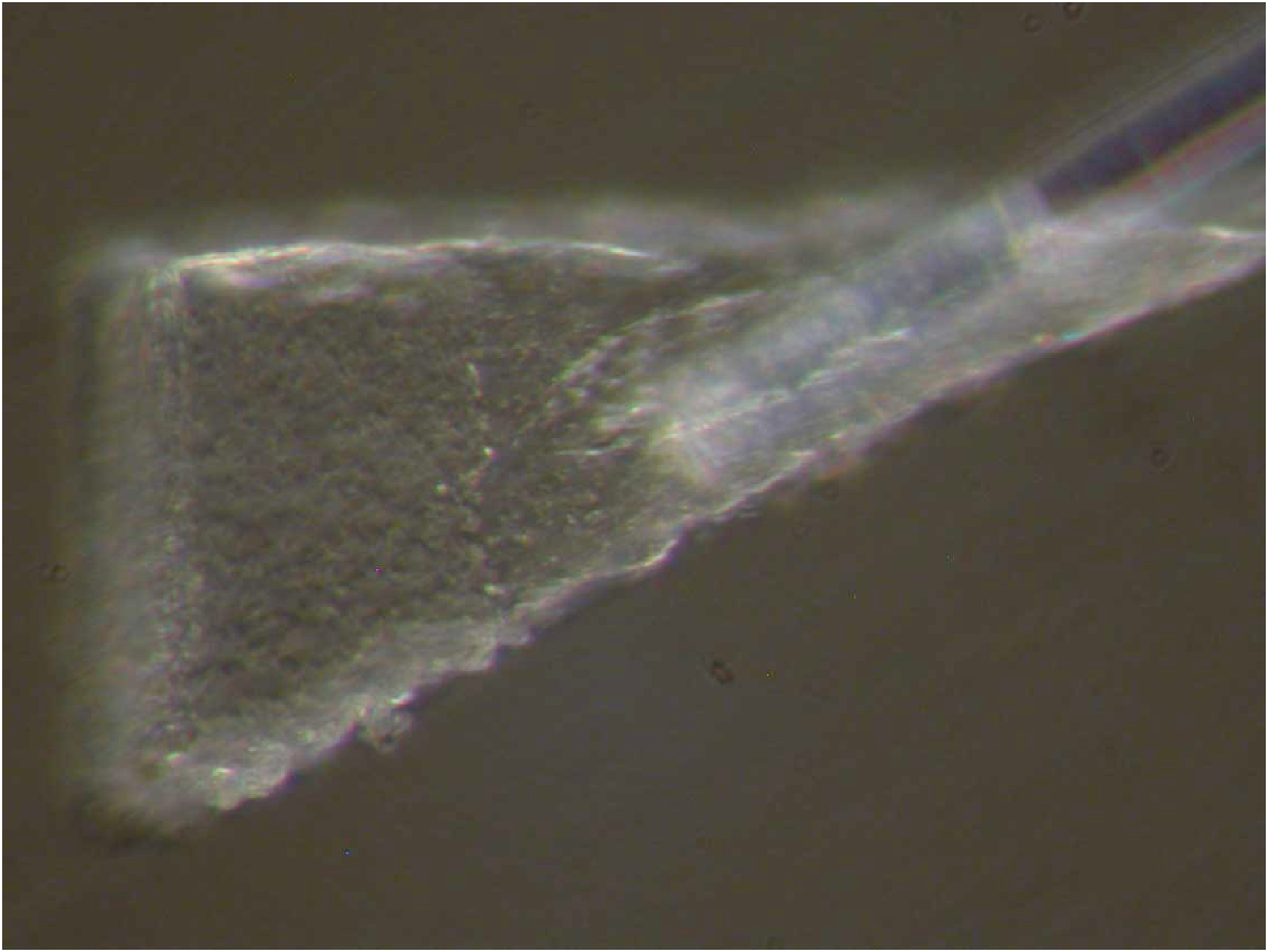}}}
\caption[dissection]{A dissection keystone  (19sep03) extracted from the ODCE collector mounted on a microforklift.  Longest dimension is 650 $\mu$m. } \label{dissection}
\end{center}
\end{figure}

For analyses in which it is desirable to expose the residue on the inside wall of the
track, it is also possible to cut vertically along the axis of the hypervelocity section of
the track.  (Fig. \ref{dissection}).   This results in a ``dissection keystone''; this approach 
has the advantage that, while it extracts half of the hypervelocity region track for analysis, it leaves
 the remainder of the track in the collector for later extraction.  
 The terminal particle and any particle residue in the track whisker cannot be divided by this method, and are either removed in dissection keystone or are left in the collector.

\begin{figure}
\begin{center}
\rotatebox{0}{\resizebox{!}{4cm}{\includegraphics*{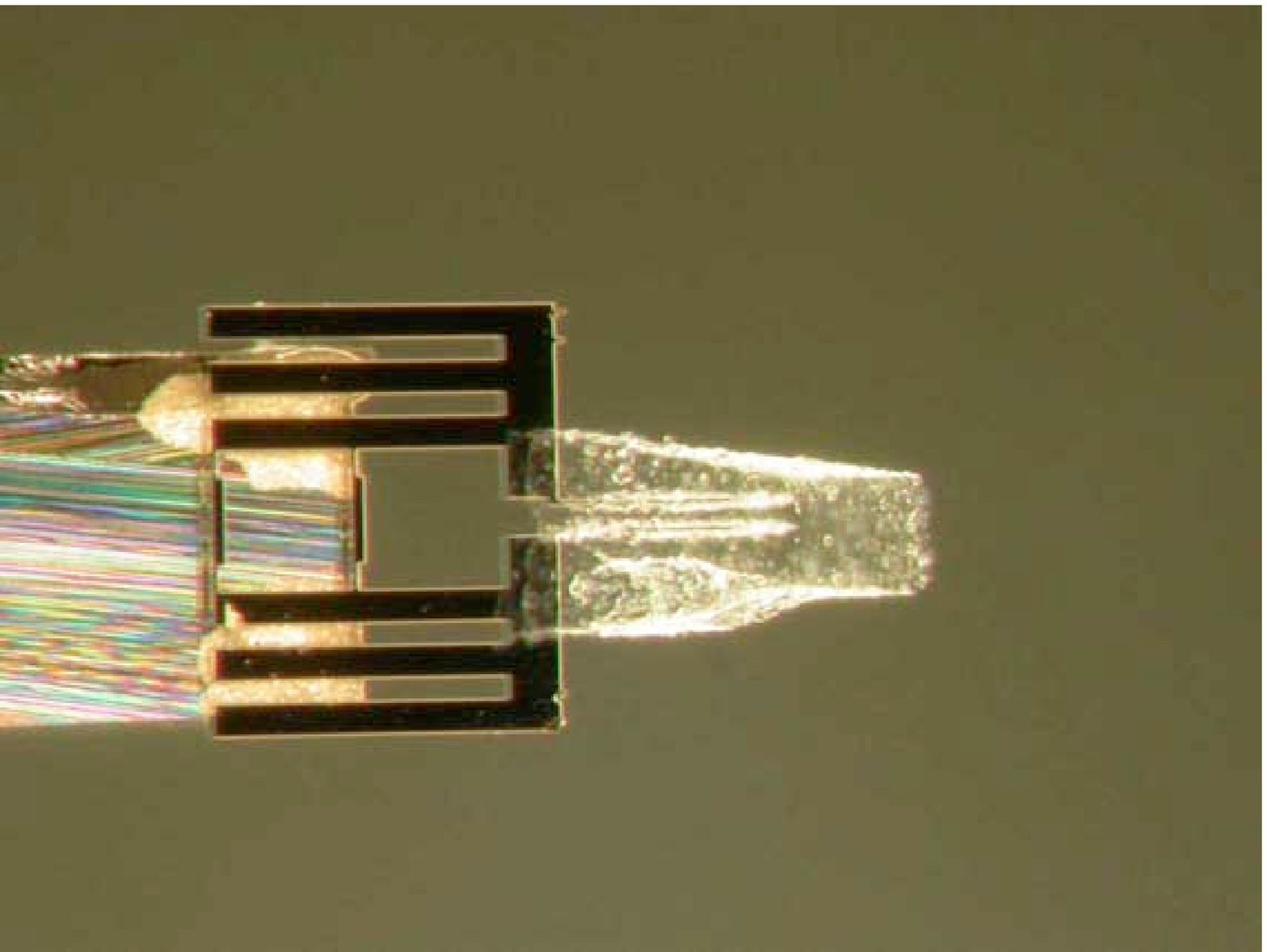}}}
\rotatebox{0}{\resizebox{!}{4cm}{\includegraphics*{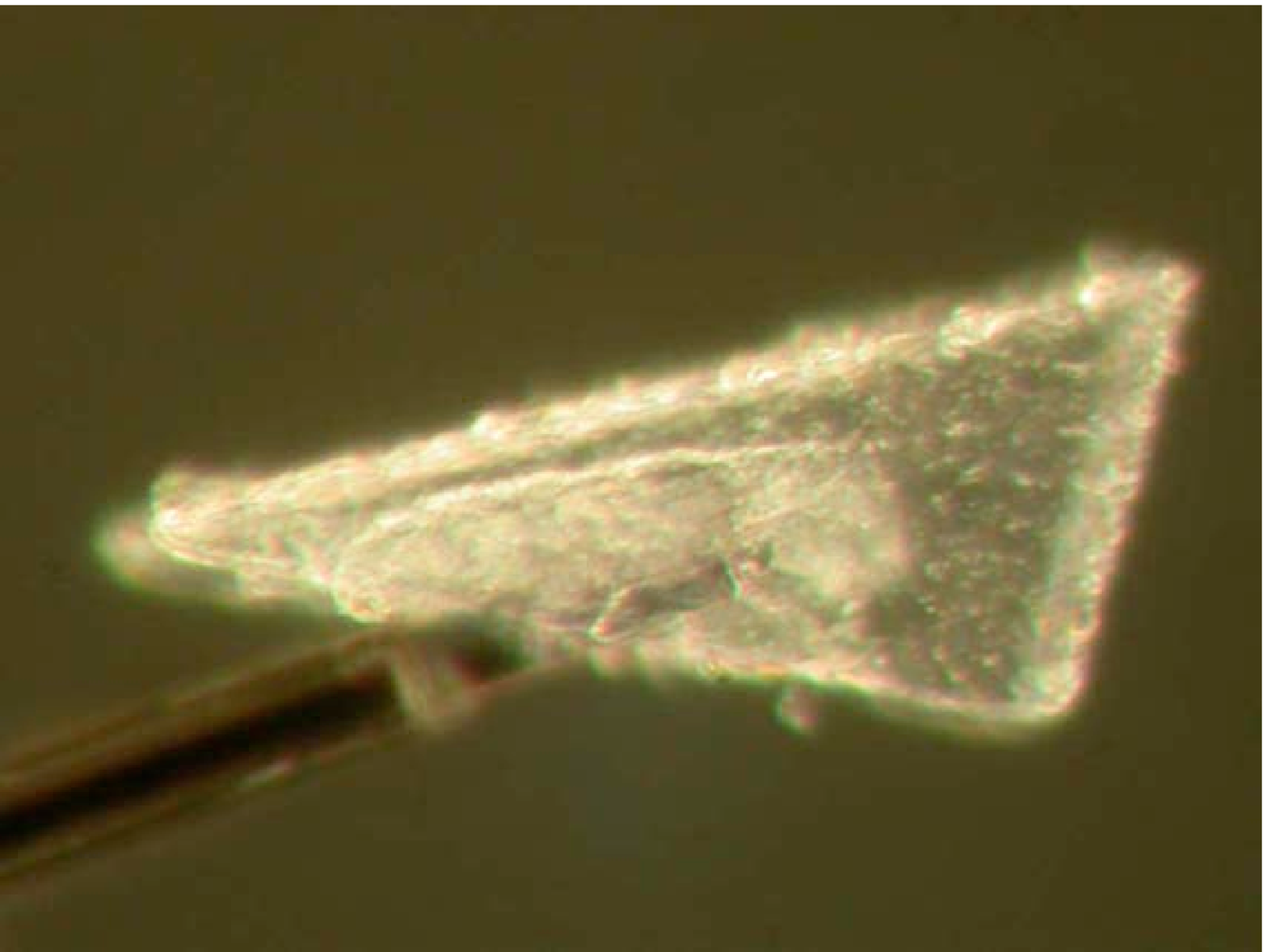}}}
\rotatebox{0}{\resizebox{!}{4cm}{\includegraphics*{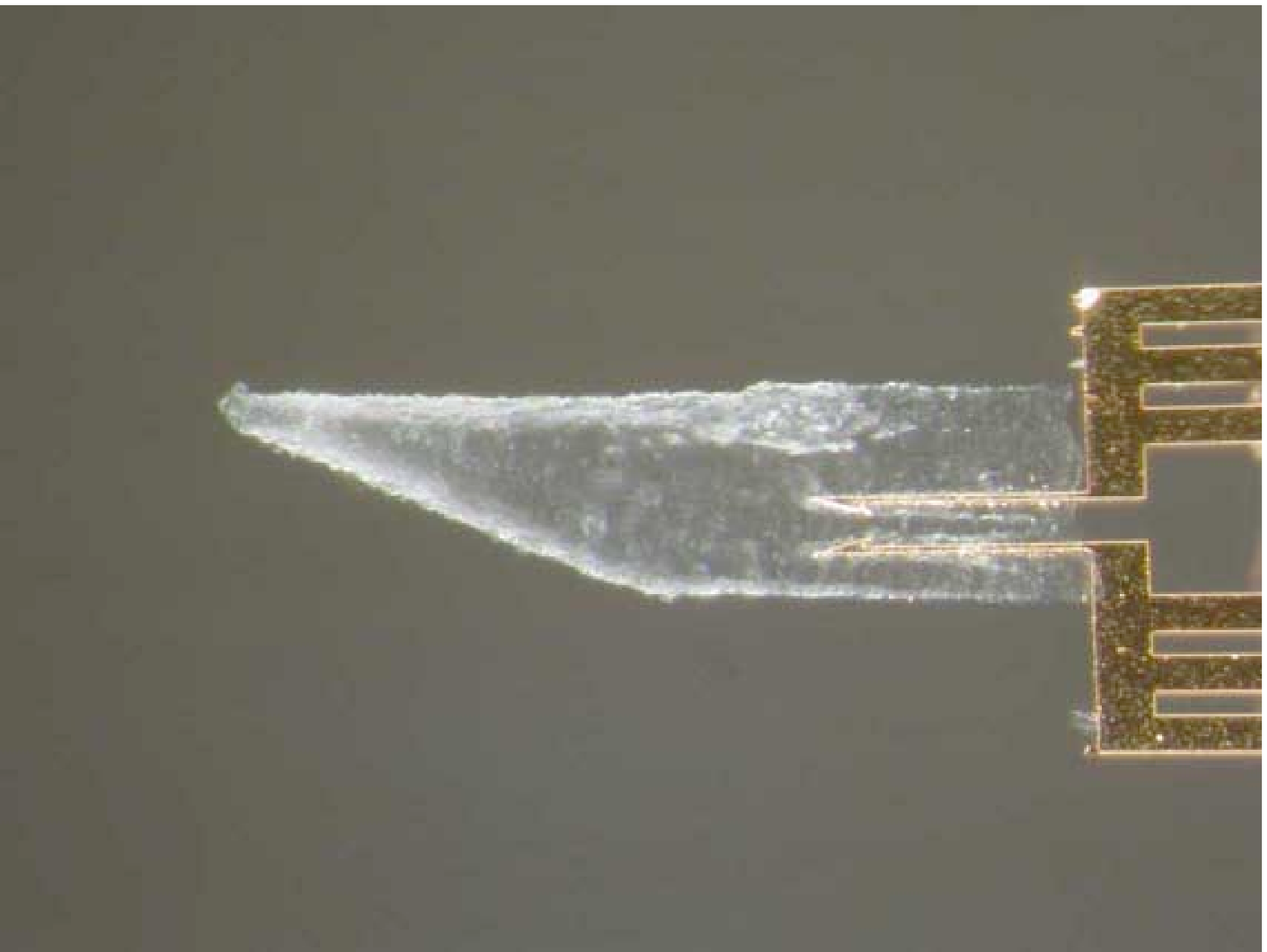}}}
\rotatebox{0}{\resizebox{!}{4cm}{\includegraphics*{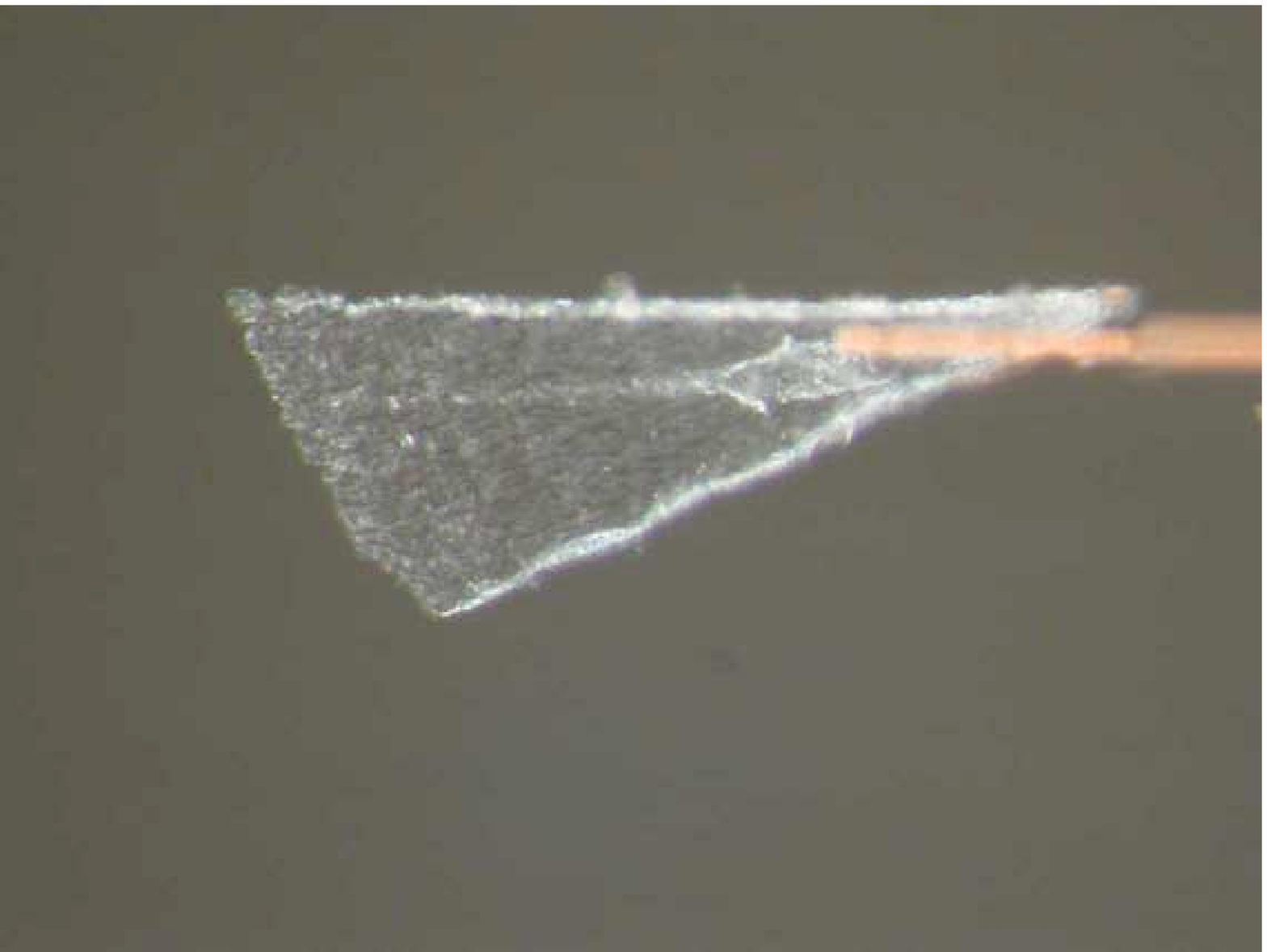}}}
\caption[keystone]{Two aerogel keystones extracted from the ODCE mounted on microforklifts:  Keystone 16jan03 (a) top view and ,
  (b) side view.  Longest dimension is 550 $\mu$m.  Keystone 24jul03 (c)  top view and (d) side view.  Longest dimension is 1100 $\mu$m. } \label{keystone}
\end{center}
\end{figure}

The extracted keystones are small,  fragile, and susceptible to loss due to even very gentle
air currents.  Working with Christopher Keller at MEMS Precision Instruments, we have developed a set of micromachined fixtures for keystones, which we call ``microforklifts".  Keystones fixed on the microforklifts (Fig. \ref{keystone}) are self-supporting, so there is no need for mounting on any additional interfering substrate that could complicate some microanalytical techniques.  
The microforklifts, in turn, are mounted on 1-mm diameter glass rods that can be readily
handled.

\noindent{\bf }

\noindent{\bf }

\begin{figure}
\begin{center}
\rotatebox{0}{\resizebox{!}{7cm}{\includegraphics*{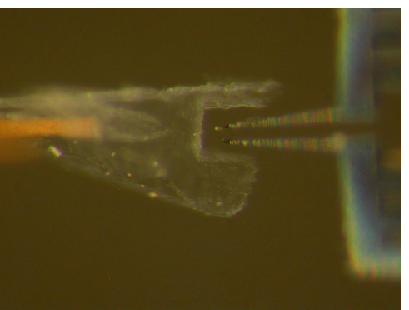}}}
\caption[extraction-from-keystone]{An extration of an individual particle from an aerogel keystone.} \label{extraction-from-keystone}
\end{center}
\end{figure}

Extraction of individual grain fragments at the track terminus or along the track may be required after the initial characterization of all preserved material within the aerogel keystone has been achieved 
(Fig. \ref{extraction-from-keystone}).  
We have demonstrated that individual particles can be extracted from keystones
using custom-developed micromachined silicon microtweezers developed by us in collaboration with MEMSPI.  These tweezers are able to routinely handle micron-scale particles.  We are continuing the development effort on these tweezers, including an embeddable, encoded tweezer than can be used for precise positioning during microtoming.  However, we repeat our cautionary statement here 
that extraction of individual, large particles may bias the analysis toward robust minerals.

\noindent{\bf }

\ \\{\bf \Large Summary of techniques and sample preparation methods} \ \\

In Table 1 we summarize  the sample preparation techniques that may be required for common analyses.

\begin{table}
\begin{tabular}{|l|l|c|c|c|c|c|c|}%
\hline\hline 
{\bf Analytical}				& {\bf Analytical}   & \multicolumn{6}{c}{\bf Sample preparation technique} \vline \\
\cline{3-8}
{\bf Goal}					& {\bf Technique} & keystone     & flattened & microtomed   & FIB-mined & microtomed/  & aerofilm \\
						&                             & (containment or& keystone& flattened &  keystone & FIB-thinned      &                 \\
						&                             &  dissection)  &               & keystone  &                        & particle        &                 \\
\hline
Elemental 				& SEM/EDX	   & R	      &  $\bullet$  &   $\bullet$      &  $\bullet$  &   $\bullet$   &   $\bullet$ \\
\cline{2-8}
Composition 				& Nuclear            &  	              &                 &                        &                     &                      &                 \\
 			                            & microprobe      & R		    &    $\bullet$ &    $\bullet$   &   $\bullet$ &   $\bullet$    &   $\bullet$ \\
 			                            & (STIM/PESA/    &                &                  &                        &                     &                      &                 \\
 			                            & \ \ RBS/PIXE)   &                &                  &                        &                     &                      &                 \\
\cline{2-8}
			                            & EMPA     	&	X	&       R           &     $\bullet$   &   $\bullet$ &   $\bullet$    &   $\bullet$ \\
\cline{2-8}
			                            & XRF ($Z \ge 6$) &	 R	&      $\bullet$ &    $\bullet$   &   $\bullet$ &   $\bullet$    &   $\bullet$ \\
\hline
Min/Pet			                   & XRD		 &	R	&    $\bullet$ &    $\bullet$   &   $\bullet$ &   $\bullet$    &   $\bullet$ \\
\cline{2-8}
				                   & Modal  		&	R	&    $\bullet$   &    X?   &  $\bullet$? &   X   &   X \\
				                   &Mineralogy 	&		&                  &                        &                     &                      &                 \\
\cline{2-8}
				                   & Tomography &	R	&                  &                        &                     &                      &                 \\
\cline{2-8}
				                   & FTIR 		&	X?	&       R           &     $\bullet$   &   $\bullet$ &   $\bullet$    &   X? \\
\cline{2-8}
				                   & Raman 		&$\bullet$ &    $\bullet$ &    $\bullet$   &   $\bullet$ &   $\bullet$    &   $\bullet$ \\
\cline{2-8}
				                   & STEM 		&	X	&       X           &            R            &           X          &            R          &        $\bullet$?         \\
\hline 
Chemistry			                   & STXM		&	X	&       X           &            R            &           X          &            R          &        $\bullet$?         \\

\cline{2-8}
				                   & FTIR		 &	X?	&       R           &     $\bullet$   &   $\bullet$ &   $\bullet$    &   X? \\
\cline{2-8}
				                   & $\mu$L$^2$MS   &     X	&       R           &     $\bullet$   &   $\bullet$ &   $\bullet$    &   $\bullet$ \\
\hline
Isotopes		                  		& SIMS		 &     X	&       R           &     $\bullet$   &   $\bullet$ &   $\bullet$    &   $\bullet$ \\
\cline{2-8}
			                  		& nanoSIMS	 &     X	&       R           &     $\bullet$   &   $\bullet$ &   $\bullet$    &   $\bullet$ \\
\cline{2-8}
			                  		& RIMS		 &     X	&       R           &     $\bullet$   &   $\bullet$ &   $\bullet$    &   $\bullet$ \\
\hline\hline
\end{tabular}
\caption{Summary of particle and impact residue sample preparation techniques for some common analytical techniques.  Symbol key:
R = Minimum requirement; $\bullet$ = Possible but not required; X = Not possible.}
\end{table}

Several  analytical techniques (e.g., nuclear microprobe and SEM/EDX, 
described in a companion paper in this series;  XRF and XRD) require no further sample preparation beyond the extraction of an event in a keystone and mounting on a microforklift.  However, other analytical techniques require further sample preparation.     We show a summary of techniques and sample
preparation methods in Table 1.

\ \\{\bf \large Analytical techniques requiring flattened keystones} \ \\

\begin{figure}
\resizebox{6.0in}{!}{\includegraphics*{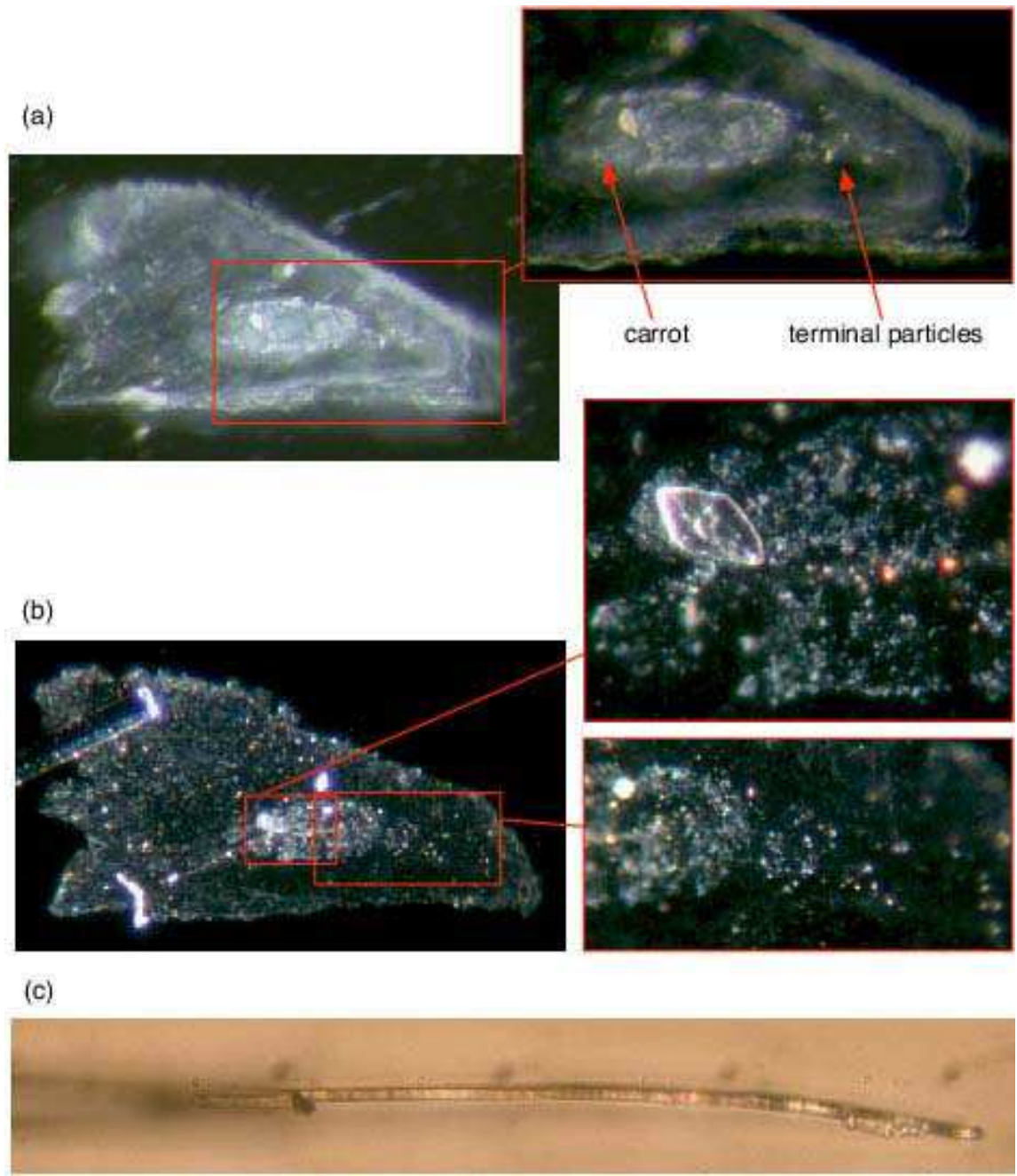} }
\caption[flattening] {Demonstration of flattening of an aerogel keystone containing a chondritic swarm impact.  The  keystone was flattened between
two glass slides coated with non-adhesive coatings of carbon and Au+Pd, respectively.   a) Keystone before flattening.
The thickest part of the keystone wedge (into the page)  is $\sim300$ $\mu$m thick, and its longest dimension is 550$\mu$m.  Multiple terminal particles
are visible in whisker tracks downstream of the carrot. 
b)  Face-on image of flattened keystone.   In addition to the terminal particles,
individual small particles distributed along the particle track can be resolved.  The residual particles exhibit a striking variety of colors, so 
appear to be heterogeneous.  
c) Edge-on mosaic image of the flattened  keystone held 
in a pair of micromachined tweezers.    The flattened keystone thickness is $6\mu$m, a factor of $\sim$50 thinner than the unflattened keystone.
The observed warpage
in this keystone is not typically observed when flattening onto polished Al metal or EMBED 812 epoxy. 
 } \label{flattening}
\end{figure}

Ionization Mass Spectrometry (SIMS, RIMS) requires a physically flat sample in order to avoid
strong electric potential errors due to the strong electric field at the sample.  Because of its extremely
short working distance, the nanoSIMS requires a particularly flat ($\le 1\mu$m) sample.
For different physical reasons, FTIR analysis also requires a flat (or dispersed) sample 
in order to avoid interference effects samples that are physically extended in height. 
We have found that aerogel keystones
can be readily flattened between glass slides without shattering.   The collapse appears to be essentially homologous:
the projected images of the keystones appear to be nearly identical before
and after flattening (Fig. \ref{flattening}).   For techniques in which material is ablated either slowly
(e.g., nanoSIMS) or not at all (EPMA), it may be desirable to prepare a flattened {\em dissection}
keystone, so that the particle residue is completely exposed for analysis.

\noindent{\bf }

\ \\{\bf \large Analysis techniques requiring thin sections} \ \\

\begin{figure}
\begin{center}
\rotatebox{0}{\resizebox{!}{7cm}{\includegraphics*{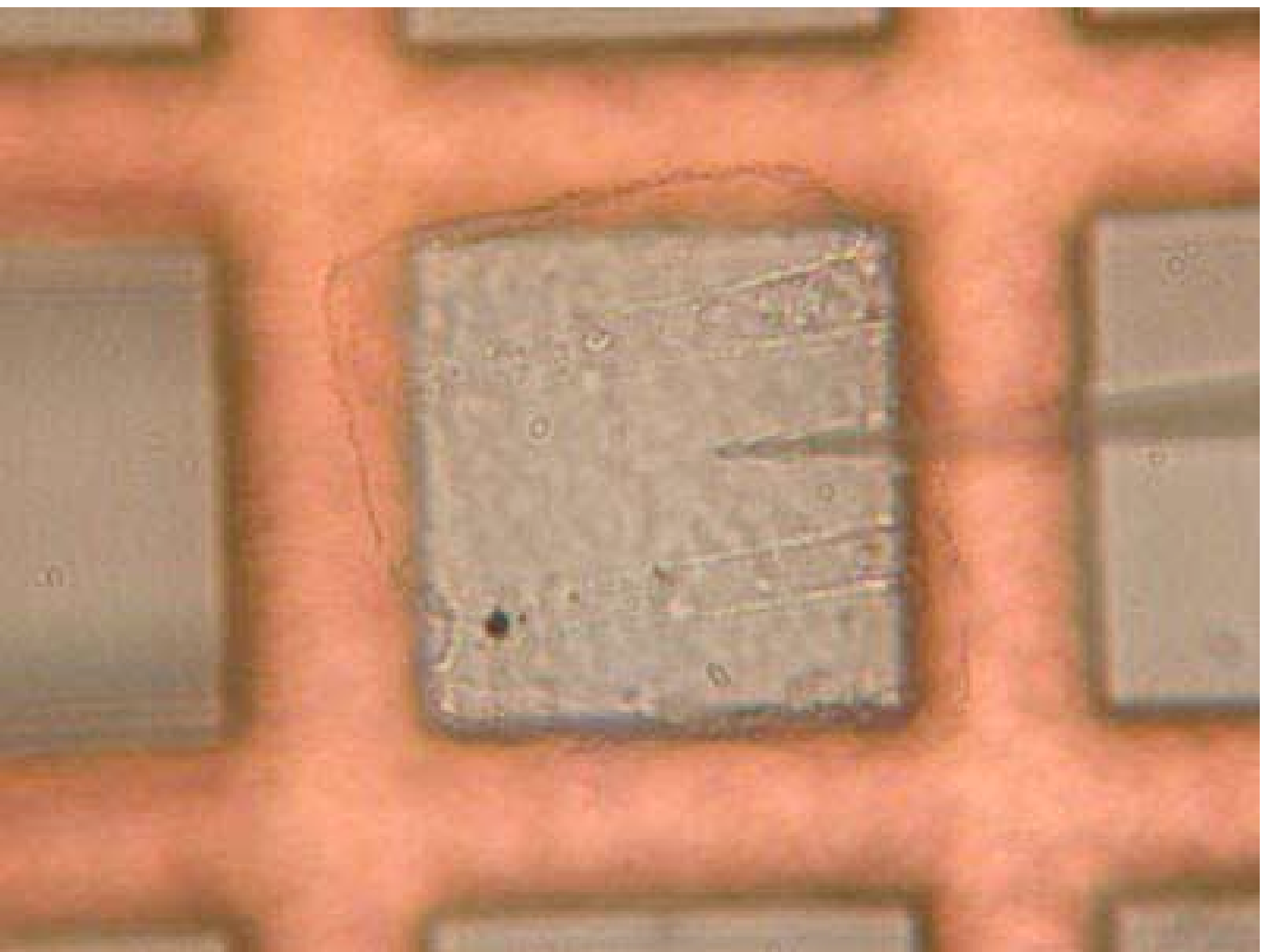}}}
\rotatebox{0}{\resizebox{!}{7cm}{\includegraphics*{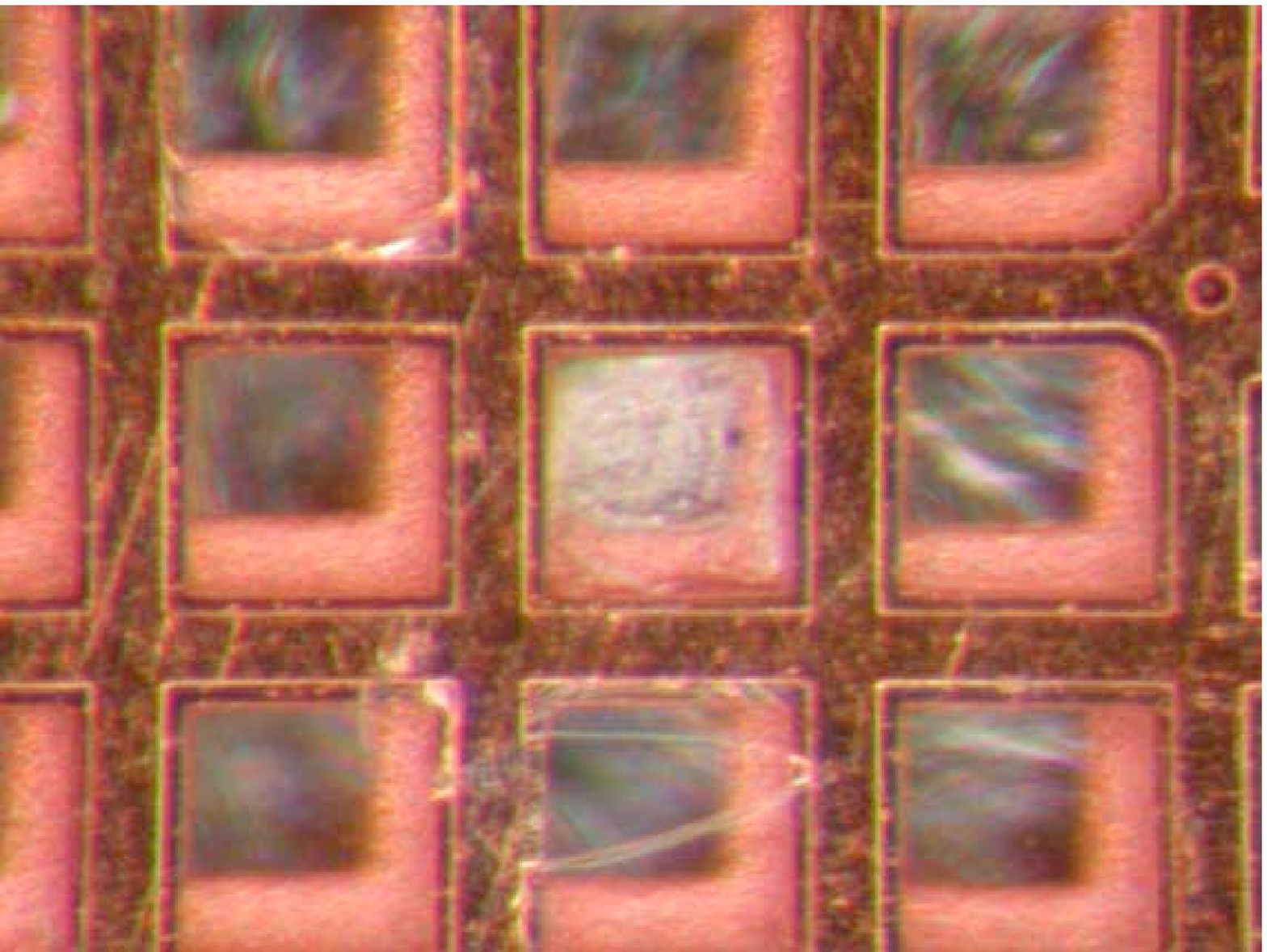}}}
\caption[aerofilm]{A 20 $\mu$m thick ``aerofilm'' placed into a 97 $\mu$m $\times$ 97 $\mu$m TEM grid.    (top) Emplacement using a glass microneedle (bottom) An aerofilm trapped between two TEM grids.  } \label{aerofilm}
\end{center}
\end{figure}

TEM analysis requires ultrathin sections, which may be prepared from aerogel keystones
in several different ways.   We have demonstrated that aerogel keystones may be flattened by a factor of $\sim10$, embedded
in epoxy (e.g., EMBED 812) and ultramicrotomed.    As a next step, we plan to develop the capability of ultramicrotomy
of keystones embedded in sulfur.  Individual grains, extracted
using actuated microtweezers, may also be embedded and microtomed.
Graham {\em et al} have demonstrated that thin sections from samples may be prepared
using FIB milling\cite{fib}.  We have demonstrated that individual thin
sections of aerogel (``aerofilms'') only 20 $\mu$m thick can be extracted from aerogel collectors
and placed into TEM grids for analysis (Fig. \ref{aerofilm}).  This may be useful
for the analysis of very small ($<100$ nm) particulates in aerogel, in which embedding may be
undesirable.

\noindent{\bf }

\ \\{\bf \large Extraction of interstellar dust impacts: Picokeystones} \ \\

\begin{figure}
\begin{center}
\rotatebox{0}{\resizebox{!}{7cm}{\includegraphics*{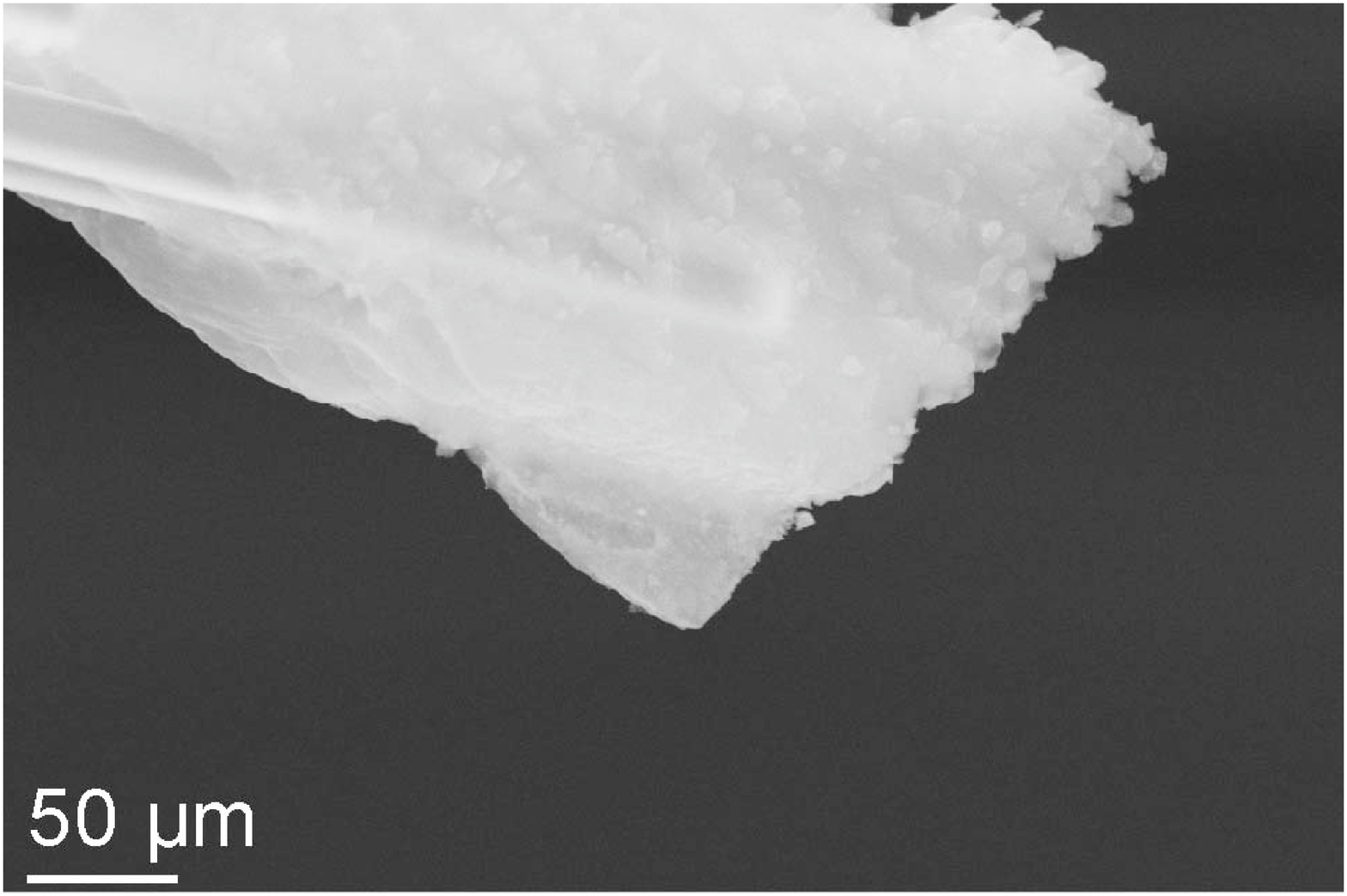}}}
\caption[picokeystone]{A secondary electron micrograph of a $\sim 30\mu$m-long track extracted in a ``picokeystone''. 
The track has approximately the length and geometry of that expected from an interstellar
dust particle captured in the Stardust interstellar collector.} \label{picokeystone}
\end{center}
\end{figure}

In addition to the cometary dust collector, the Stardust mission also carries 
an interstellar dust collector.    Landgraf \cite{landgraf} has estimated  that Stardust will collect
40 interstellar dust particles of mass 1 pg or greater.  Because of the steeply falling
mass spectrum, most of these particles will be sub-micron in size.  The identification, extraction
and analysis of these particles will be extremely challenging.  We have taken the first
steps in the extraction of very small impact tracks that may be similar to those of IS grains.  
First, we machine away material on either side of a wedge-shaped volume
of aerogel that contains the small track.  Next, we extract the track from
the collector in an ordinary keystone.  The result is a stacked keystone (a ``picokeystone'', since the projectile probably has a mass on the
picogram scale) in
which the large keystone serves as a carrier for the small keystone containing the track.
(Fig. \ref{picokeystone}).

\noindent{\bf }

\ \\{\bf \large Extraction of particle impacts from crowded fields} \ \\

It is possible that the number of particles collected by Stardust will greatly exceed the
Stardust goal, and that, if the collectors survive, we will be faced with the delightful 
challenge of harvesting a bonanza of captured
cometary material.  In this case, the collectors may be optically opaque near
the surface, preventing the initial imaging and targeting of individual grains.  
A possible approach to the extraction of these particles would be to extract particles
in a series of layered keystones.  First, a small keystone would be extracted, which would
contain stopping small particle residues and only portions of the hypervelocity sections of
tracks of larger particles.  If this extraction removes the opaque surface layer, deeper particles might then be imaged and targeted for individual extraction.  
If not, successively deeper and deeper keystones could
be extracted below the first, each containing the terminal particles and residues of larger and larger impacting particles.

\ \\{\bf \Large Current development efforts} \ \\

Here we have described only some first steps in the development of sample preparation techniques
for the analysis of the precious returned samples from Stardust.     We are refining our extraction techniques to improve the accuracy and speed, and are currently developing techniques for embedding particles using encoded, microtomable microtweezer tips, so that particles can be readily located inside opaque embedding media.  

In this development effort we have been guided by our own at best semi-educated guesses 
regarding analytical requirements.  Future directions in sample
preparation should be dictated  by the requirements of the analytical techniques, but these
have not yet been adequately defined for Stardust samples.   It is critical that these requirements be defined as soon as possible so that the community
can adequately prepare for the return of the Stardust mission.

A major roadblock to the development of these requirements
is the lack of understanding of the capture process in aerogel.  Some   unanswered questions are:

\begin{itemize}%
\item What  thermal and pressure profile is experienced by particles in the 1-30 $\mu$m range as they are captured in aerogel at 6 km sec$^{-1}$?  What alteration does this thermal and pressure pulse cause in inorganic minerals and compounds of astrophysical interest?  
\item What trace element contaminants are present in the aerogel, and to 
what extent will these contaminants limit the analysis of captured particles?
\item To what extent will organics native to the aerogel limit the analysis of organics native to the captured particles?  What new organics, both in the aerogel and in the projectiles, 
might be formed during the capture process?  Are new compounds synthesized by reactions
between projectile and aerogel organics?
\end{itemize}%

The answers to these questions are vitally important for the success of Stardust, and because
they will serve to define the requirements for analytical techniques, they will drive the particle
extraction and sample preparation development efforts in the future.  

\bibliographystyle{unsrt}

\typeout{References here}

\noindent\textbf{\large Acknowledgments}\\
This work at UCB was supported by NASA Grant NAG5-10411.  The work carried out by G. Graham, J. Bradley, S. Bajt, P.G. Grant and G. Bench was performed under the auspices of the U.S. Department of Energy, National Nuclear Security Administration by the University of California, Lawrence Livermore National Laboratory under contract No. W-7405-Eng-48.  The work carried out by S. Brennan and P. Pianetta was supported by the Department of Energy, Office of Basic Energy Sciences, under contract DE-AC03-76SF00515.

\noindent Correspondence and requests for materials should be addressed to AJW. (e-mail: westphal@ssl.berkeley.edu).

\end{document}